\documentclass{rQUF2e}
\usepackage{etex}
\usepackage{epstopdf}
\usepackage{subfigure}
\usepackage{lineno,hyperref}
\usepackage{amsmath,mathtools,amssymb,amsthm}
\usepackage{pstricks}
\usepackage{pstcol}
\usepackage{indentfirst}
\usepackage{psfrag}
\usepackage{tabularx}
\usepackage{pstricks-add}
\usepackage{dsfont}
\usepackage{graphicx}
\usepackage{url}
\usepackage[]{algorithm2e}
\usepackage{bigints}
\usepackage{pdflscape}
\usepackage{longtable,tabu}
\usepackage{ltablex,booktabs}
\usepackage{caption} 
\usepackage[font={small}]{caption}
\def\bibsection{\section*{References}}
\newtheorem{proposition}{Proposition}
\newtheorem{theorem}[proposition]{Theorem}

\newtheorem{remark}[proposition]{Remark}

\newtheorem{q}[proposition]{Question}

\begin{document}
\title{Singular Fourier-Pad\'e Series Expansion of European Option Prices}

\author{Tat Lung (Ron) Chan$^{\ast}$$\dag$\thanks{$^\ast$Corresponding author.
Email: t.l.chan@uel.ac.uk} \\
\affil{$\dag$University of East London, Water Lane, Stratford, UK, E15 4LZ\\
\received{v1.1 released May 2016} }}

\maketitle

\begin{abstract}
We apply a new numerical method, the singular Fourier-Pad\'e (SFP) method invented by \cite{Dris_Ben:2001,Dris_Ben:2011}, to price European-type options in L\'evy and affine processes. The motivation behind this application is to reduce the inefficiency of current Fourier techniques when they are used to approximate piecewise continuous (non-smooth) probability density functions. When techniques such as fast Fourier transforms and Fourier series are applied to price and hedge options with non-smooth probability density functions, they cause the Gibbs phenomenon; accordingly, the techniques converge slowly for density functions with jumps in value or derivatives. This seriously adversely affects the efficiency and accuracy of these techniques. In this paper, we derive pricing formulae and their option Greeks using the SFP method to resolve the Gibbs phenomenon and restore the global spectral convergence rate. Moreover, we show that our method requires a small number of terms to yield fast error convergence, and it is able to accurately price any European-type option deep in/out of the money and with very long/short maturities. Furthermore, we conduct an error-bound analysis of the SFP method in option pricing. This new method performs favourably in numerical experiments compared with existing techniques.
\end{abstract}

\section{Introduction}\label{sec:intro}
The application of Fourier techniques to pricing and hedging financial derivatives has  flourished in computational and mathematical finance over the last decade because most of the underlying processes driving derivatives have a characteristic function (i.e., the Fourier transform of the probability density function (PDF)). In general, two  Fourier techniques are applied to option pricing: fast Fourier transforms (FFTs) and Fourier series, such as Fourier cosine (COS) series. To explain the first technique, FFTs in option pricing, we first consider a European option pricing formula that is a risk-neutral expectation of the discounted payoff of the option driven by an underlying process. This formation naturally implies the integration of the discounted payoff and the PDF of the underlying process. As the underlying process has a characteristic function, one can apply an inverse continuous Fourier transform to recover the PDF, discretise the continuous Fourier integral and use an FFT algorithm that computes discrete Fourier transforms in $O(N \log_2(N))$ operations to obtain the option prices \citep[cf.][]{Car_Mad:1999, Lew:2001, Lip:2002, Cho:2004, Itk:2005, Lor_Fan:2008}. The main disadvantage of using these methods is that they require thousands of grid points and considerable  computational time to reach an acceptable level of accuracy. Another approach is to use Fourier series to expand a PDF in terms of a partial sum of orthogonal basis functions. The basis function can be complex an exponential, e.g., \cite{Chan:2016}, or COS, e.g. \cite{Fan_Oos:2009a}, function. Each coefficient in the series is approximated via the characteristic function of the PDF. By integrating each basis function with the discounted payoff function of the option, we can obtain an option price. This technique originated in \cite{Fan_Oos:2009a} to price European-style options with COS basis functions. The success of Fourier COS series in European option pricing has lead to the rapid expansion of pricing and hedging options with early-exercise features and exotic options, like Asian, multi-asset or barrier options \citep[e.g.,][]{Lee_Oos:2008, Fan_Oos:2009b, Fan_Oos:2011, Zha_Oos:2013}. Compare with FFTs, the advantage of using Fourier series for option pricing is that they can achieve a global spectral (exponential) convergence rate and require fewer summation terms as long as the governing PDF is sufficiently smooth. However, using any type of Fourier series to represent a $\mathbf{C}^\nu$\footnote{A vector space in which functions are $\nu$ continuously differentiable.} piecewise continuous (non-smooth) function\footnote{A function is called piecewise continuous on an interval if the function is made up of a finite number of $\nu$ times differentiable continuous pieces.}, e.g., a non-smooth PDF, is notoriously fraught. 
 Discontinuities cause the Gibbs phenomenon, which has two important consequences for the Fourier partial sum of length $N$:
\begin{enumerate}
\item failure to converge at the jump, and
\item pointwise convergence elsewhere at the rate $O(N^1)$. 
\end{enumerate}
More generally, if the function $f$ and its derivatives up to order $\nu-1$ are continuous but $f^{(\nu)}$ is discontinuous (i.e., $f$ has a jump of order $\nu$), then the global convergence rate is $O(N^{-\nu})$. The impact of the Gibbs phenomenon can lead to inaccurate pricing and hedging when the approximate option prices are generated via  FFT or Fourier series methods at or around the jumps.

\cite{Ruj_Oos:2013} attempt to resolve the Gibbs phenomenon and improve the algebraic-index convergence rate of the plain COS method by adding filters. 
They call this the filter COS method. In their paper, they intensively test six different filters, e.g., the Fej\'er filter and the exponential filter, and discover that the exponential filter gives a better algebraic-index convergence rate for pricing European options. The method is implemented to represent the non-smooth PDF of the Variance Gamma (VG) model. Although they provide some improvements over the current COS method, filtering only works well away from the peak of the VG PDF; at the peak, the approximation performs worse. The method also requires almost a thousand COS partial summation terms to reach acceptable accuracy when the PDF is non-smooth.

According \cite{Gra_Oos:2013,Gra_Oos:2016}, another disadvantage of using the COS method is that the COS exhibits periodicity in the vicinity of the integration boundaries, and long-maturity option round-off errors may accumulate near the domain boundaries. A series of papers of using wavelets, e.g., B-spline and Shannon wavelets, have addressed this problem. Wavelet methods are similar to Fourier series methods; they use wavelets  to represent a PDF and then integrate the discounted payoff function with wavelet basis functions to reach the closed-form representation of an option pricing formula. Compared with the COS method, the wavelet methods are flexible and accurate for pricing options with long maturities. However, the choice of the wavelet basis functions determines whether the method can achieve a spectral convergence rate. In some cases, the B-spline and Haar wavelets cannot achieve spectral convergence (cf. Table 1 in \citealp{Gra_Oos:2013} and \citealp{Gra_Oos:2016}). Although Ortiz-Gracia and Oosterlee use Shannon wavelets (Shannon Wavelets Inverse Fourier Technique (SWIFT)) to correct this problem and achieve spectral convergence in some cases where the PDF is smooth (cf. Table 1 and Table 4 in \cite{Gra_Oos:2016}), the accuracy of SWIFT heavily relies on a scale parameter. The scale parameter is determined by the speed of decay of a characteristic function, which in turn, determines the accuracy of the approximation of the corresponding PDF. Although, in general, \cite{Gra_Oos:2016} consider 0 or 1 the best choice of the scale, there is no theoretical reason for this choice, and different scale parameters are chosen for different PDFs throughout their paper.

Finally, in our numerical tests, the accuracy of the COS method is very sensitive to small and large option prices. The COS method tends to be less accurate when we measure small option prices (cf. Table \ref{table:VG_Smooth_SFP1} in Section \ref{sec:results}).

To circumvent the aforementioned disadvantages of FFT, Fourier series and wavelet methods, we propose a novel method, the singular Fourier-Pad\'e (SFP) method, which exhibits the following characteristics: 
\begin{enumerate}
\item global spectral convergence rate for piecewise continuous PDFs,
\item fast error convergence with fewer partial summation terms required,
\item accurately prices any European-type option with the features of deep in/out of the money and very long/short maturities, 
\item consistently accurate for approximating large or small option prices throughout, and
\item does not require a scale parameter to adjust its accuracy. 
\end{enumerate}

Why do we choose the SFP method? Compared with the SFP method, the Fourier-Pad\'e technique is not a better choice. The Fourier-Pad\'e technique \citep[cf.][]{Chi_Com:1981,Sma_Cha:1988,Gee:1995} is a famous technique for approximating a non-smooth function such that spectral convergence  can be achieved away from the jumps and convergence is not globally degraded. Nevertheless, \cite{Dris_Ben:2001,Dris_Ben:2011} note that the fundamental limitation of the Fourier-Pad\'e method is that the use of poles to approximate branch cuts is inefficient, and logarithmic singularities translated by the jumps on the unit circle are very difficult for the Pad\'e approximants to simulate. Accordingly, the method fails to converge at the jump. To address this problem, \cite{Dris_Ben:2001,Dris_Ben:2011} add proper logarithmic branch singularity terms into the Fourier-Pad\'e approximation process. They call this method the singular Fourier-Pad\'e method. In all numerical tests \citep[cf.][]{Dris_Ben:2001,Dris_Ben:2011}, the SFP can accelerate error convergence to the true solutions for analytic periodic functions or non-smooth functions, and the global effects of the Gibbs phenomenon due to a jump are largely eliminated. If the function is very difficult, when jump locations are known in advance, the method still appears to converge globally and exponentially and offer 4--6 digits of accuracy at the jumps. Most importantly, in addition to comparing the Fourier-Pad\'e technique, Driscoll and Fornberg compare the SFP method with other two famous methods of overcoming the Gibbs phenomenon--the Gegenbauer method \citep{Got_Shu:1997}, a method to project the Fourier partial sum onto a space spanned by a Gegenbauer polynomial, and singularity removal \citep{Eck:1997}, a method of singularity removal for functions with jump discontinuities. In the numerical tests  comparing these three methods \citep[cf.][]{Dris_Ben:2001,Dris_Ben:2011}, the SFP method yields better global spectral convergence and faster error convergence than do the other methods. Based on all these benefits of the SFP method, we chose it over the other available numerical methods.
 
The remainder of this paper is structured as follows. Section \ref{sec:intro} provides an introduction. Section \ref{sec:SFP} describes the SFP method. Section \ref{sec:stocModels} introduces the financial stochastic models we examine in this paper. Section \ref{sec:option} describes and proves the formulation of the SFP option pricing/Greeks formulae for different styles of European options. In Section \ref{sec:SFP_algorithm}, we describe the SFP algorithm and the Fourier-Pad\'e method to find the jump locations on a non-smooth function. Section \ref{sec:trunc} describes the choice of truncated integration intervals. Section \ref{sec:error} provides the error analysis of the SFP method in option pricing and hedging. Section \ref{sec:results} discusses, analyses and compares the numerical results of the SFP method with those of other numerical methods. Finally, we conclude and discuss possible future developments in Section \ref{sec:conclusion}.

\section{Singular Fourier-Pad\'e Interpretation and Correction of the Gibbs Phenomenon}\label{sec:SFP}
If we consider a function $f$ with a formal power series representation $\sum_{k=0}^\infty b_k x^k,$ and a rational function defined by $R_{N,M}=P_N/Q_M,$ where $P_N$ and $Q_M$ are the polynomials of 
\begin{align}\label{eqn:FP_1}
P_N(x)=\sum_{n=0}^N p_{n} x^{n}\hbox{ and } Q_M(x)=\sum_{m=0}^M q_{m} x^{m},
\end{align}
respectively, then we say that $R_{N,M}=P_N/Q_M$ is the \textsl{(linear) Pad\'e approximant} of order (N, M) of the formal series satisfying the condition  
\begin{align}\label{eqn:FP_2}
\left(\sum_{n=0}^N p_{n} x^{n}\right)-\left(\sum_{m=0}^M q_{m} x^{m}\right)\left( \sum^{M+N}_{k=0} b_k x^k\right)&=\mathcal{O}(x^{N+M+1}).
\end{align}
Here, $f$ is approximated by $\sum^{M+N}_{k=0} b_k x^k,$ To obtain the approximant $R(N, M),$ we simply calculate the coefficients of polynomials $P_N$ and $Q_M$ by solving a system of linear equations. To obtain $\{q_m\}_{m=0}^M,$ we first normalise  $q_0 = 1$ to ensure that the system is well-determined and has a unique solution in (\ref{eqn:FP_2}). Then, we consider the coefficients for $x^{N+1},\ldots , x^{M+N},$ and we can yield a Toeplitz*\footnote{A Toeplitz matrix or diagonal-constant matrix is an invertible matrix in which each descending diagonal from left to right is constant.} linear system:
\begin{align}
\begin{bmatrix} 
b_{N+1} &  b_{N} & b_{N-1} & \cdots & b_{N+1-M}\\ 
b_{N+2} &  b_{N+1} & b_{N} & \ddots & b_{N+2-M}\\ 
\vdots&\ddots &\ddots &\ddots &\vdots\\
b_{N+M} & \cdots &b_{N+2} & b_{N+1} & b_{N}
\end{bmatrix}
\begin{bmatrix}
q_0\\q_1\\ \vdots\\ q_M
\end{bmatrix}
=0.
\end{align}
Once $\{q_m\}_{m=0}^M$  is known, $\{p_n\}_{n=0}^N$ is found through the terms of
order N and less in (\ref{eqn:FP_2}). This yields $\mathbf{p} = B\mathbf{q}$, where $b_{ij} = b_{i-j}$. For example, if
$N=M,$ one obtains
\begin{align}
\begin{bmatrix}
p_0\\p_1\\ \vdots\\ p_N
\end{bmatrix}=\begin{bmatrix} 
b_{0} &  &   &   \\ 
b_{1} &  b_{0} &  & \\ 
\vdots&\ddots &\ddots &\\
b_{N} & \cdots & b_{1} & b_{0}
\end{bmatrix}
\begin{bmatrix}
q_0\\q_1\\ \vdots\\ q_M
\end{bmatrix}.
\end{align}
Computing Pad\'e approximants through linear algebra, as we have done here, is simple but not necessarily the most efficient or stable numerical method \citep[cf.][]{Dris_Ben:2001,Dris_Ben:2011}.

Now, suppose that $f$ be a piecewise analytic function defined on the interval $[-\pi, \pi),$ with $s$ jump locations in $f$ at $t = \zeta_s\in[\pi, \pi)$, $s=1,\ldots,S.$ We define a jump as an actual discontinuous point on $f$ or a discontinuous point appearing after its derivatives. Then the complex Fourier series (CFS) representation is given by 
\begin{align}
f(t)=\sum_{k=-\infty}^{\infty} b_k e^{ikt},\quad\hbox{with}\quad b_k=\frac{1}{2\pi}\int_{-\pi}^{\pi} f(t)e^{-ikt} dt. 
\end{align}
The transformation $z=e^{it}$, which maps the interval $[\pi,\pi)$ onto the unit circle in the complex plane, transforms the Fourier series into the following Laurent series in $z$, which can be split into
\begin{align}
f(z)&=\sum_{k=-\infty}^{\infty} b_k e^{ikt}=\sum_{k=0}^\infty{\vphantom{\sum}}' b_k z^k + \sum_{k=0}^\infty{\vphantom{\sum}}'b_{-k} z^{-k}\nonumber\\
&=f^+(z)+f^{-}(z^{-1}), 
\end{align}
where the prime sums indicate that the zeroth term should be halved. The Fourier-Pad\'e approximation of $f_{\pm}$ is comprised of polynomials 
\begin{align}\label{eqn:FP}
P_N^\pm(z)=Q_M^\pm(z)f^\pm(z)+\mathcal{O}(z^{N+M+1}),	\quad z\rightarrow 0.
\end{align}
The resulting approximant is then defined as
\begin{align}
{P^+_N(z)\over Q^+_M(z)}+{P^-_N(z^{-1})\over Q^-_M(z^{-1})}.
\end{align}
However, \cite{Dris_Ben:2001,Dris_Ben:2011} note that this approximant does not reproduce very well at/around the jump locations of the function and make the approximation inaccurate. Therefore, they suggest that every jump in value of $f$ at $t=\zeta$ can be attributed to a logarithm of the form
\begin{align}
\log\left(1-{z\over e^{i\zeta}}\right).
\end{align}
This logarithmic singularity in $f_\pm$, which is difficult for the Pad\'e approximant to simulate, can be exploited to enhance the approximation process. This is the rationale behind the SFP method introduced
in \cite{Dris_Ben:2001,Dris_Ben:2011}. We modify the Fourier-Pad\'e approximant (\ref{eqn:FP}) to obtain the following condition:
\begin{align}
P_N^\pm(z)+\mathcal{L}=Q_M^\pm(z)f^\pm(z)+\mathcal{O}(z^{U+1}),
\end{align}
where
\begin{align}
\mathcal{L}=\sum_{s=1}^S L^{\pm}_s(z)\log\left(1-{z\over e^{i\zeta_s}}\right)
\end{align}
for some polynomials $L_s, s = 1,\ldots, S$, and $U$ is determined by S and the degrees of $P_N$, $Q_M$ and $L_s$.

If we extend the SFP method to support any piecewise analytic real function $f$ in a finite interval $[a, b]$ with a set of jump locations $\{\zeta_s\}_{s=1}^S\in[a,b]$ appearing in $f$, the CFS representation of the function is defined as 
\begin{eqnarray}
f(x)=\mathfrak{Re}\left[\sum_{k=-\infty}^{\infty} b_k e^{i\frac{2\pi}{b-a}k x}\right],\,\hbox{with}\,\, b_k=\frac{1}{b-a}\int_{a}^{b} f(x)e^{-i\frac{ 2\pi }{b-a}kx} dx. \label{FourierS}
\end{eqnarray}
Here, $\mathfrak{Re}$ represents the real part of the function. 
As we focus on approximating a real function, we can further obtain
\begin{eqnarray}
f(x)=\mathfrak{Re}\left[2\sum_{k=1}^{\infty} b_k e^{i\frac{2\pi}{b-a}kx}+b_0\right]\label{FourierS_Final}.
\end{eqnarray}
Based on this representation, we denote $z$ as $\exp\left({i\frac{2\pi}{b-a}x}\right)$ and then transform $f$ into a truncated power series of $f_1$ equal to  
\begin{eqnarray}\label{eqn:CFS_ztransform}
\mathfrak{Re}\left[2\sum_{k=1}^{U} b_k z^k+b_0\right].
\end{eqnarray}
The transformation $z=\exp\left({i\frac{2\pi}{b-a}x}\right)$ also suggests that the jump location $\zeta$  translates into 
\begin{eqnarray}
\log\left(1-{z\over \varepsilon}\right), \hbox{ where } \varepsilon=e^{i\frac{2\pi}{b-a}\zeta}\label{eqn:log_jump}
\end{eqnarray}
in $f_1.$ Finally, applying (\ref{eqn:CFS_ztransform}) and (\ref{eqn:log_jump}), we can reach the final form of the SFP approximant given by
\begin{align}\label{eqn:SFP_1}
P_N^+(z)+\sum_{s=1}^S L_{N_s}^+(z)\log\left(1-z/\varepsilon_s \right)=\left(2\sum_{k=1}^{U} b_k z^k+b_0\right)Q_M^+(z)+\mathcal{O}(z^{U+1}),
\end{align}
where
%
\begin{eqnarray}
\begin{array}{rclrcl}
P_N^+(z)&=&\sum_{n=0}^N p_{n} z^{n},&Q_M^+(z)&=&\sum_{m=0}^M q_{m} z^{m}\neq 0,\\
L_{N_s}^+(z)&=&\sum_{n_s=0}^{N_s}l_{n_s} z^{n_s},& s&=&1,\ldots, S,\\
U&=&N+M+S+\sum_{s=1}^S N_s.&&&
\end{array}
\end{eqnarray}
Once we have determined the unknown coefficients of $\{p_n\}_{n=0}^N,$ $\{q_m\}_{m=0}^M$ and $\{l_{n_s}\}_{n_s=0}^{N_s}$ (cf. Section \ref{sec:SFP_algorithm}), the SFP representation of $f(x)$ can be formulated as
\begin{align}
f(x)=\mathfrak{Re}\left({P_N^+(z)+\sum_{s=1}^S L_{N_s}^+(z)\log\left(1-z/\varepsilon_s \right) \over Q_M^+(z)}\right),\hbox{ with } z=e^{i\frac{2\pi}{b-a}x}.
\end{align}

\section{Stochastic Models of the Asset Dynamics}\label{sec:stocModels}
We assume frictionless financial equity markets and no arbitrage, and take as given an equivalent martingale measure $\mathds{Q}$ chosen by the market. All stochastic processes defined in the following are assumed to live on the complete filtered probability space $(\Omega, \mathcal{F}, \{\mathcal{F}_t\}_{t\geq 0},\mathds{Q}).$  Standard references for exponential L\'evy processes can be found in \citet{Sch:2003} and \citet{Con_Tan:2004}; for affine processes, see \citet{Duf_Sch:2003}.

\subsection{Exponential L\'evy Processes}\label{eqn:stock_Lvy}
The stock price process $(S_t)_{t\geq0}$ under $\mathds{Q}$ driven by an exponential L\'evy process can be defined:
\begin{align}\label{eqn:stock_Lvy}
S_T&=S_te^{(r-q)(T-t)+X_T-X_t+\omega(T-t)}\\
&=S_te^{(r-q+\omega)(T-t)+X_{T-t}},
\end{align}
where $X_T,$ $X_t$ and $X_T-X_t$ are all L\'evy processes. As L\'evy processes have  independent stationary increments, we can say that $X_T-X_t=X_{T-t}.$ Throughout the paper, $r\geq 0$ and $q\geq 0$ denote the constant risk-free interest rate and the constant dividend yield, respectively; $S_t$ represents the known stock price at time $t$; and $S_T$ represents the random stock price at time $T$. The condition that $(S_Te^{-(r-q)(T-t)})_{t\geq 0}$ is a martingale will be guaranteed by an appropriate choice of the mean-correcting compensator $\omega$ as follows:
\begin{align}
\omega={1\over T-t}\mathbb{E}\left(e^{X_{T-t}}\right),
\end{align}
where $\mathbb{E}\left(e^{X_{T-t}}\right)$ is assumed to be finite for all $0\leq t\leq T$. Given a L\'evy process $(X_{T-t})_{t\geq 0}$, define the corresponding
characteristic function as follows:
\begin{align}
\varphi(u)=\mathbb{E}[e^{iu(X_{T-t})}]=e^{(T-t)\phi(u)},\quad u\in \mathbb{R}.
\end{align}
Here, $\phi(u)$ is a continuous function with the L\'evy-Khintchine representation given by 
$$\phi(u)={1\over 2}A^2u -i\gamma u + \int_{-\infty}^{\infty}(1-e^{iu\chi}+iu\chi\mathds{1}_{\vert \chi \vert\geq 1}) \nu(\mathrm{d} \chi), \quad\chi\in X_{T-t},$$
with characteristic triplet $(A,\gamma, \nu).$ There is substantial consideration of exponential L\'evy processes in stock process modelling. Due to space limitations, we cannot show all the processes in this paper; rather, we selectively investigate the processes due to their popularity or their numerical implementation challenge. However, this does not mean that our method only works on the processes we selected. As the paper evolves, we can see that as long as a process has a characteristic function, the option price and its risk driven by the process can be properly approximated via our method.
\subsection{Exponential Affine Processes}
An exponential affine characteristic function is given by 
\begin{align}
\varphi(u)=\mathbb{E}[e^{iuX_t}]=\exp(C(u,t)+X_0D(u,t)).
\end{align}
Here, the function $C(u,t)$ is fully characterised by $dC(u,t)/dt=D(u,t)$ and the function $D(u,t)$ satisfies a Riccati equation \citep{Duf_Sch:2003}. More precisely, we can find functions $C(u,t)$ and $D(u,t)$ with $C(u,t)=0$ and $D(u,t)=u,$ such that
\begin{align}
M_t=\exp(C(u,T-t)+X_tD(u,T-t))
\end{align} is a martingale process. In the family of affine processes, we focus on investigating the Heston model in this paper. The stochastic differential equation (SDE) of the Heston model is written as 
\begin{align}
dS_t&=(r-q)S_tdt+\sqrt{y_t}S_tdW_{1,t},\\
dy_t&=\lambda(\bar{y}-y_t)dt+\eta\sqrt{y_t}dW_{2,t},
\end{align}
where $L_t$ and $y_t$ denote the stochastic log-asset price variable and the variance of the asset price process, respectively. In this process, the speed of mean reversion $\lambda$, the mean level of variance $\bar{y}$ and the volatility of volatility $\eta$ are constant values greater than or equal to zero. Additionally, the Brownian motions $W_{1,t}$ and $W_{2,t}$ are correlated with the correlation coefficient $\rho_s$, and $\omega=\mathbb{E}[e^{i(-1i)X_t}]$ is the mean-correcting compensator. The model characteristic function fits in the general affine characteristic function framework and is given by
\begin{align}\label{eqn:hestonChar}
\varphi(u)&=\exp\bigg(iu\left((r-q)t+\omega\right)+\frac{y_0}{\eta^2}\left(\frac{1-e^{Et }}{1-Fe^{E t }}(\lambda-i\rho_s\eta u-E)\right) +\nonumber\\
\,&\frac{\lambda\bar{y}}{\eta^2}\left( t (\lambda-i\rho_s\eta u-E)-2\log(\frac{1-Fe^{-E t }}{1-F})\right)\bigg), \,\, u\in\mathbb{R},\\
\intertext{with}
E=&\sqrt{\left(\lambda-i \rho_s \eta u \right)+(u^2+iz)\eta^2},\,\nonumber\\
F=&(\lambda-i\rho_s\eta u-E)/(\lambda-i\rho_s\eta u+E).\nonumber
\end{align}
This characteristic function is uniquely specified because we take $\sqrt{(x+y i)}$ such that its real part is nonnegative, and we restrict the complex logarithm to its principal branch. 

In this case, as \citet{Lor_Kah:2010} prove, the resulting characteristic function is the correct one for all complex numbers $z$ in the analytic strip of the characteristic function. In the SDE, we have two possible conditions with respect to $\lambda,$ $\bar{y}$ and $\eta$:
\begin{align}
2\lambda\bar{y}\geq\eta^2\label{eqn:con1},\\
2\lambda\bar{y}<\eta^2\label{eqn:con2}.
\end{align}
The model satisfies the Feller property if (\ref{eqn:con1}) holds; otherwise, (\ref{eqn:con2}) holds. If a process fulfils the property, the process never hits zero; conversely, if it does not, the process can reach 0. Condition (\ref{eqn:con2}) is a very important property for the Heston SDEs because they can only have a unique solution when we specify a boundary condition at 0. In mathematical finance, the chosen boundary condition is that the process remains at 0. We define this as an absorbing boundary condition. When the process reaches 0 and is allowed to leave 0, we call it a reflecting boundary. These two boundary conditions are crucial for pricing early-exercise options, including American options and barrier options.

\section{Singular Fourier-Pad\'e Representation of European Option Prices and their Option Greeks}\label{sec:option}
\subsection{Option Pricing Formulae}
In this section, we derive closed-form formulae for European-style options using the SFP method. 
Considering the PDF $f$ of a stochastic process, the current log-price $x := \log S,$  the strike price of $K$ and maturity $T \geq t,$ we can express the option price $V(x,K,t)$ starting at time $t$ with its contingent claim paying out $G(S_T, K)$ as follows:
\begin{align}\label{eqn:GEquation_1}
V(x,K, t)&=e^{-r (T - t ) } \mathbb{E } (G(S_T, K) \vert S_t = e^x )\\
&=e^{-r (T - t ) } \mathbb{E } (G(S_te^{X_T-X_t}, K))\\
&=e^{-r(T-t)}\bigintssss_{-\infty}^{\infty} G(e^{x + \chi},K) f(z) dz,\quad \chi \in X_T-X_t.
\end{align}
Furthermore, if we choose an interval $[c,d]$ satisfying  the condition  
\begin{align}\label{eqn:charAppx}
\int_{c}^{d} f(\chi) e^{iu\chi} \rm{d}\chi\approx\mathbb{E}[e^{iu(X_T-X_t)}]=\varphi(u), 
\end{align}
where $\varphi(u)$ is a characteristic function of $X_T-X_t,$ we can approximate the pricing formula:
\begin{eqnarray}\label{eqn:GEquation_2}
V(x,K, t)\approx e^{-r(T-t)}\int_{c}^{d} G(e^{x + \chi},K) f(\chi) \mathrm{d}\chi.
\end{eqnarray}


\begin{theorem}
\label{EuroCallOption}
When a dividend-paying risky asset price process $(S_t)_{t\geq 0}$ with a traceable, analytical characteristic function $\varphi(\cdot)$ has a current asset price $e^x=S,$ risk-free interest rate $r$  and compounded continuous dividend $q,$ the SFP pricing formula of a \textbf{European vanilla call option} driven by this process with maturity $T$ and strike price $K$ is
\begin{align}
V(x,K,t)&=e^{-r(T-t)}\mathfrak{Re}\left[{ P_N^+(z)+\sum_{s=1}^S L^+_{N_s}(z)\log\left(1-z/\varepsilon_s \right) \over Q^+_M(z)} \right], \label{eqn:EuroCall_SFP}
\end{align}
with
\begin{align}
\begin{array}{rclrcl}
z&=&e^{i\frac{2\pi}{d-c}(-x+\log K)}&\varepsilon_s&=&e^{i\frac{2\pi}{d-c}\zeta_s}\\
P_N^+(z)&=&\sum_{n=0}^N p_{n} z^{n},&Q_M^+(z)&=&\sum_{m=0}^M q_{m} z^{m}\neq 0,\\
L_{N_s}^+(z)&=&\sum_{n_s=0}^{N_s}l_{n_s} z^{n_s},& s&=&1,\ldots, S.\\
\end{array}
\end{align}
Here, $\zeta_s$ is the jump in $V(x,K,t).$ 

\end{theorem}
\noindent\textbf{Proof:}
First, the payoff function of a European call option is $G(e^{x+\chi},K)=\max\left(e^{x+\chi}-K,0\right)$ in  (\ref{eqn:GEquation_2}). To conform to the SFP approximation framework, we first transform the payoff: 
 \begin{align}
\max\left(e^{x+\chi}-K,0\right)=K\max\left(e^{x+\chi-\log K}-1,0\right). 
 \end{align}
Accordingly, by replacing  $x+\chi-\log K$ with $y$, we have a new form of $V(x,K,t)$ denoted as
\begin{align}
e^{-r(T-t)}\int_{-\infty}^{\infty} K\max(e^{y}-1,0) f\left(y-x+\log K\right) \mathrm{d}y.
\end{align}
As we intend to make the SFP method more efficient, we define a truncated computational interval $[c,d]$ (cf. Section \ref{sec:trunc}), which satisfies condition $(\ref{eqn:charAppx}),$  to replace $[-\infty,\infty].$ Then, $V(x,K,t)$ is reformulated as
\begin{align}\label{eqn:EuroV_1}
\quad&e^{-r(T-t)}\int_{c}^{d}K\max(e^{y}-1,0) f\left(y-x+\log K\right) \mathrm{d}y\nonumber\\
=&e^{-r(T-t)}\int_{0}^{d} K(e^{y}-1) f\left(y-x+\log K\right) \mathrm{d}y.
\end{align}
Using the Fourier transform shift theorem and the CFS expansion shown in (\ref{FourierS_Final}),  we express $f\left(y-x+\log K\right)$ as
 \begin{align}\label{eqn:CFS_PDF}
\mathfrak{Re}\left[2\sum_{k=1}^{\infty} b_k e^{i\frac{2\pi}{b-a}ky}+b_0\right],
\end{align}
where
\begin{align}
b_k=\frac{1}{d-c}\int_{c}^{d} f(y)e^{-i\frac{ 2\pi }{d-c}ky} \mathrm{d}y\left(e^{i\frac{2\pi}{d-c}k\left(-x+\log K\right)}\right)\quad\hbox{and}\quad b_0=\frac{1}{d-c}\int_{c}^{d} f(y)\mathrm{d}y.
\end{align}
We substitute (\ref{eqn:CFS_PDF}) into (\ref{eqn:EuroV_1}) and apply Fubini's theorem;  $V(x,K,t) $ can be computed as 
\begin{align}\label{eqn:EuroCall_CFS_temp}
&\quad e^{-r(T-t)}\int_{0}^{d} K(e^{y}-1)\mathfrak{Re}\left[2\sum_{k=1}^{\infty} b_k e^{i\frac{2\pi}{d-c}ky}+b_0\right] \mathrm{d}y\nonumber\\
&=e^{-r(T-t)}\mathfrak{Re}\Bigg[2\sum_{k=1}^{\infty}\frac{1}{d-c}\int_{c}^{d} f(y)e^{-i\frac{ 2\pi }{d-c}ky} \mathrm{d}y\left(e^{i\frac{2\pi}{d-c}k\left(-x+\log K\right)}\right)\int_{0}^{d} K(e^{y}-1) e^{i\frac{2\pi}{d-c}ky}\mathrm{d}y+\nonumber\\
&\quad \frac{1}{d-c}\int_{c}^{d} f(y)\mathrm{d}y\int_{0}^{d} K(e^{y}-1)\mathrm{d}y\Bigg).
\end{align}
In the equation above, basic calculus implies that 
\begin{align}\label{eqn:G_k}
\int_{0}^{d} K(e^{y}-1) e^{i\frac{2\pi}{d-c}ky}\mathrm{d}y&=\frac{K(d-c)}{i 2\pi k+(d-c)}\left(e^{(i\frac{2\pi}{d-c}k+1)d}-1\right)-\frac{K(d-c)}{i 2\pi k}\left(e^{(i\frac{2\pi}{d-c}k)d}-1\right),\\
\int_{0}^{d} K(e^{y}-1) \mathrm{d}y&=K(e^d-1)-K d.
\end{align}
In the meantime, because of condition (\ref{eqn:charAppx}), we can also see that 
\begin{align}\label{eqn:B_k}
\int_{c}^{d} f(y)e^{-i\frac{ 2\pi }{d-c}ky} \mathrm{d}y\approx\varphi\left(-\frac{2\pi}{d-c}k\right)\quad\hbox{and}\quad \int_{c}^{d} f(y)\mathrm{d}y\approx\varphi(0)=1. 
\end{align}
For the sake of simplicity in (\ref{eqn:EuroCall_CFS_temp}), we set 
\begin{align}\label{eqn:BkGk}
\begin{array}{rclrcl}
\widehat{B}_k&=&\frac{1}{d-c}\varphi\left(-\frac{2\pi}{d-c}k\right),&\widehat{B}_0&=&{1\over d-c}\varphi(0)={1\over d-c},\\
\widehat{G}_k&=&\frac{K(d-c)}{i 2\pi k+(d-c)}\left(e^{(i\frac{2\pi}{d-c}k+1)d}-1\right)-\frac{K(d-c)}{i 2\pi k}\left(e^{(i\frac{2\pi}{d-c}k)d}-1\right),& \widehat{G}_0&=&K(e^d-1)-K d\\
\end{array}
\end{align}
to obtain a simplified form: 
 \begin{align}\label{eqn:EuroCall_CFS}
V(x,K,t)&=e^{-r(T-t)}\mathfrak{Re}\left[2\sum_{k=1}^{\infty} \widehat{B}_k\widehat{G}_k  e^{i\frac{2\pi}{d-c}k\left(-x+\log K\right)} + \widehat{B}_0\widehat{G}_0\right].
\end{align}


To express our final pricing formula with the SFP representation, we approximate 
\begin{align}\label{eqn:temp1}
2\sum_{k=1}^{\infty} \widehat{B}_k\widehat{G}_k  e^{i\frac{2\pi}{d-c}k\left(-x+\log K\right)} + \widehat{B}_0\widehat{G}_0
\end{align}
in (\ref{eqn:EuroCall_CFS}) with (\ref{eqn:SFP_1}). To do so, we first replace $-x+\log K$ with a variable $y_1$ in (\ref{eqn:temp1}), and then, this allows us to set  $\exp\left({i\frac{2\pi}{d-c} y_1}\right)$ equal to $z.$ The transformation $z=\exp\left({i\frac{2\pi}{d-c} y_1}\right)$ maps the interval $[c,d]$ onto the unit circle in $z.$ This change also transforms the jumps $\zeta$ along $V(x,K,t)$ into $z$ with the form of $\varepsilon=\exp\left(i\frac{2\pi}{d-c}\zeta\right).$ Finally, expressing (\ref{eqn:temp1}) with a new variable of $z,$ we have 
\begin{align}\label{eqn:temp2}
2\sum_{k=1}^{\infty} \widehat{B}_k\widehat{G}_k z^k + \widehat{B}_0\widehat{G}_0.
\end{align}
Substituting the equation above with $f_1(z)$ in (\ref{eqn:SFP_1}), we obtain the approximant  given by
\begin{align}\label{eqn:EuroCall_SPF1}
P_N^+(z)+\sum_{s=1}^S L^+_{N_s}(z)\log\left(1-z/\varepsilon_s \right)=\left(2\sum_{k=1}^{U} \widehat{B}_k\widehat{G}_k z^k + \widehat{B}_0\widehat{G}_0\right)Q^+_M(z)+\mathcal{O}(z^{U+1})
\end{align}
\begin{align}
\begin{array}{rclrcl}
P_N^+(z)&=&\sum_{n=0}^N p_{n} z^{n},&Q_M^+(z)&=&\sum_{m=0}^M q_{m} z^{m}\neq 0,\\
L_{N_s}^+(z)&=&\sum_{n_s=0}^{N_s}l_{n_s} z^{n_s},& s&=&1,\ldots, S,\\
\varepsilon_s&=&e^{i\frac{2\pi}{d-c}\zeta_s},&U&=&N+M+\sum_{s=1}^S N_s.
\end{array}
\end{align}
Once we can determine the unknown coefficients of $\{p_n\}_{n=0}^N,$ $\{q_m\}_{m=0}^M$ and $\{l_{n_s}\}_{n_s=0}^{N_s}$ in ({\ref{eqn:EuroCall_SPF1}) via the algorithm shown in Section \ref{sec:SFP_algorithm} and replace
$$2\sum_{k=1}^{\infty} \widehat{B}_k\widehat{G}_k  e^{i\frac{2\pi}{d-c}k\left(-x+\log K\right)} + \widehat{B}_0\widehat{G}_0$$
with
$${P_N^+(z)+\sum_{s=1}^S L^+_{N_s}(z)\log\left(1-z/\varepsilon_s \right) \over Q^+_M(z)},\quad z=\exp\left({i\frac{2\pi}{d-c}(-x+\log K)}\right)$$
in ({\ref{eqn:EuroCall_CFS}), we reach our first SFP pricing formula for European vanilla call options, as in (\ref{eqn:EuroCall_SFP}). Q.E.D.

We can apply the same technique to seek the SFP pricing formula of vanilla put options. To achieve this, we first transform the put payoff function $G(e^{x+\chi},K)=\max\left(K-e^{x+\chi},0\right)$ into 
 \begin{align}
-K\max\left(e^{x+\chi-\log K}-1,0\right). 
 \end{align}
Then, we follow the proof of Theorem \ref{EuroCallOption} to reach the CFS representation of $V(x,K,t)$ given by
\begin{align}
e^{-r(T-t)}\mathfrak{Re}\left[2\sum_{k=1}^{\infty} \widehat{B}_k \widehat{G}_k e^{i\frac{2\pi}{d-c}k\left(-x+\log K\right)}+ \widehat{B}_0\widehat{G}_0 \right],\label{eqn:EuroPut_CFS}
\end{align}
with
\begin{align}
\widehat{G}_k&=\int_{c}^{0} -K(e^{y}-1) e^{i\frac{2\pi}{d-c}ky}\mathrm{d}y=\frac{K(d-c)}{i 2\pi k+(d-c)}\left(e^{(i\frac{2\pi}{d-c}k+1)c}-1\right)-\frac{K(d-c)}{i 2\pi k}\left(e^{(i\frac{2\pi}{d-c}k)c}-1\right),\\
\widehat{G}_0&=\int_{c}^{0} -K(e^{y}-1) \mathrm{d}y=K(e^c-1)-K c.
\end{align}
In the same manner shown in the proof of Theorem \ref{EuroCallOption}, we approximate the CFS expansion in (\ref{eqn:EuroPut_CFS}) with the SFP approximant (\ref{eqn:SFP_1}). We can therefore yield the same expression of (\ref{eqn:EuroCall_SFP}) with different coefficient values of $\{p_n\}_{n=0}^N,$ $\{q_m\}_{m=0}^M$ and $\{l_{n_s}\}_{n_s=0}^{N_s}.$

From the vanilla call and put pricing formulae, we can see that they share the same format as the SFP approximant of (\ref{eqn:EuroCall_SFP}). It is no different for other European-type options discussed in this paper. The only difference is that the values of  $\{p_n\}_{n=0}^N,$ $\{q_m\}_{m=0}^M$ and $\{l_{n_s}\}_{n_s=0}^{N_s}$ in the SFP formula of each option are completely different from one another. This is attributed to the jumps $\zeta$ in each option and their payoff functions, which lead to different Fourier transformations of $\widehat{G}_k$ and $\widehat{G}_0$ in (\ref{eqn:EuroCall_CFS}). For the reader's information, we enumerate all the options we investigate in this paper and their corresponding payoff functions $G(e^{x+\chi},K)$ and Fourier transformations $G_k$ and $G_0$ in Tables \ref{table:CFSPayoff1} and \ref{table:CFSPayoff2}.

\subsection{Option Greeks}
Now, we turn our attention to deriving the option Greeks.  Although accurately valuing financial claims plays a key role in financial modelling, the risk management (hedging) of these derivative instruments is equally important. Financial institutions manage option risk when they sell options to their clients through the analysis of the Greeks, which are defined as the sensitivities of the option price to single-unit changes in the values of state variables or parameters. Such sensitivities represent the different dimensions of the risk associated with an option. In this paper, we will focus on deriving three option Greeks---Delta, Gamma, and Vega. 
Delta, $\Delta,$ is defined as the rate of change in the option value with respect to changes in the underlying asset price; Gamma, $\Gamma$  is the rate of change of $\Delta$ with respect to changes in the underlying price; and finally, Vega is the measurement of an option's sensitivity to changes in the volatility of the underlying asset price. In general, volatility measures the amount and speed at which the price moves up and down, and it is often based on changes in the recent, historical prices of a trading instrument. Other Greeks, such as Theta, can be derived in a similar fashion; however, depending on the characteristic function, the derivation expression might be rather lengthy. We omit them here, as many terms are repeated. 

Delta is the first derivative of the value of $V$ of the option with respect to the underlying instrument price S. Hence, differentiating the CFS expansion of $V$ (\ref{eqn:EuroCall_CFS}) with respect to $S,$ we have 
\begin{align}\label{eqn:CFS_delta}
\Delta_t={\partial V(x,K,t) \over \partial S}&={\partial V(x,K,t) \over \partial x}{\partial x\over \partial S}\nonumber\\
&=e^{-r(T-t)-x}\Bigg(\mathfrak{Re}\Bigg[2\sum_{k=1}^{\infty}\left(-i\frac{2\pi}{d-c}k\right)\widehat{B}_k\widehat{G}_ke^{i\frac{2\pi}{d-c}k(-x+\log K)}\Bigg]\Bigg).
\end{align} 
In a similar fashion, we can obtain $\Gamma_t$ by differentiating $\Delta_t$ with respect to $S$ such that 
\begin{align}\label{eqn:CFS_gamma}
\Gamma_t={\partial^2 V(x,K,t) \over \partial S^2}={\partial \Delta_t \over \partial S}={\partial \Delta_t \over \partial x}{\partial x \over \partial S},
\end{align}
and eventually,
\begin{align}
\Gamma_t&=e^{-r(T-t)-2x}\Bigg(\mathfrak{Re}\Bigg[2\sum_{k=1}^{\infty}\left(i\frac{2\pi}{d-c}k\right)\left(i\frac{2\pi}{d-c}k+1\right)\widehat{B}_k\widehat{G}_ke^{i\frac{2\pi}{d-c}k(-x+\log K)}\Bigg]\Bigg).
\end{align} 
It is also easy to obtain the formula for Vega, ${\partial V\over \partial y_t},$ where $y_t$ is the initial value of the volatility at time $t.$ For example, for the Heston model, as $y_0$ is the initial value of the volatility in (\ref{eqn:hestonChar}), we derive Vega as follows:
\begin{align}
{\partial V(x,K,y_0, t)\over \partial y_0} &=e^{-r(T-t)}\Bigg(\mathfrak{Re}\Bigg[2\sum_{k=1}^{\infty}{\partial \widehat{B}_k\over \partial y_0}\widehat{G}_ke^{i\frac{2\pi}{d-c}k(-x+\log K)}\Bigg]\Bigg),
\end{align} 
with 
\begin{align}
{\partial \widehat{B}_k\over \partial y_0}={\partial \varphi(-\frac{2\pi}{d-c}k, y_0)\over \partial y_0},
\end{align} 
where $\varphi$ contains the parameter $y_0.$

To obtain our first SFP representation of $\Delta,$ we first let $y_1=-x+\log K,$  $z=\exp\left(i{2\pi\over d-c}y_1\right)$ and then transform all the jumps $\zeta$ in $\Delta_t$ into $\varepsilon=\exp\left(i{2\pi\over d-c} \zeta\right)$ in (\ref{eqn:CFS_delta}). Accordingly, this transforms the CFS representation in (\ref{eqn:CFS_delta}) into the form 
\begin{align}
f_1(z)=2\sum_{k=1}^{U}\left(-i\frac{2\pi}{d-c}k\right)\widehat{B}_k\widehat{G}_kz^k.
\end{align}
Based on the equation above, using (\ref{eqn:SFP_1}), we can eventually obtain the SFP approximant given by
\begin{align}
P_N^+(z)+\sum_{s=1}^S L^+_{N_s}(z)\log\left(1-z/\varepsilon_s \right)=f_1(z)Q^+_M(z)+\mathcal{O}(z^{U+1}).
\end{align}
Applying the approximation algorithm in Section \ref{sec:SFP_algorithm}  to determine the coefficients of $P_N^+,$ $Q^+_M,$ and $L^+_{N_s},$ we can obtain the SPF formula for $\Delta_t$ with the form  
\begin{align}
e^{-r(T-t)-x}\mathfrak{Re}\left({P_N^+(z)+\sum_{s=1}^S L^+_{N_s}(z)\log\left(1-z/\varepsilon_s \right)\over Q^+_M(z)}\right). 
\end{align} 
To determine the SFP approximant of $\Gamma_t$ and Vega, we follow the same idea of approximating $\Delta_t$ but replace $f_1(z)$ for 
\begin{align}
2\sum_{k=1}^{U}\left(i\frac{2\pi}{d-c}k\right)\left(i\frac{2\pi}{d-c}k+1\right)\widehat{B}_k\widehat{G}_kz^k\quad \mbox{and}\quad 2\sum_{k=1}^{\infty}{\partial \widehat{B}_k\over \partial y_0}\widehat{G}_kz^k.
\end{align}

\begin{table}
\small
\caption{Payoff functions and their transforms for a variety of financial contingency claims}  
\centering 
\resizebox{\textwidth}{!}{\begin{tabular}{|ccc|} 
\hline
Financial Contingency Claim&Payoff Function&Transformed Payoff Function\\[10pt]
&$G(S_T,K)$&$G(e^{x+\chi},K)$\\[10pt]
\hline
\rule{0pt}{4ex} 
Call&$\max(S_T-K,0)$&$K\max(e^{x+\chi-\log K}-1,0)$\\[10pt]
Put&$\max(K-S_T,0)$&$-K\max(e^{x+\chi-\log K}-1,0)$\\[10pt]
Covered Call&$\min(S_T,K)$&$K\min(e^{x+\chi-\log K}-1,0)+K$\\[10pt]
Cash-or-Nothing Call&$\mathds{1}_{S_T\geq K}$&$\mathds{1}_{e^{x+\chi-\log K}\geq 1}$\\[10pt]
Cash-or-Nothing Put&$\mathds{1}_{S_T\leq K}$&$\mathds{1}_{e^{x+\chi-\log K}\leq 1}$\\[10pt]
Asset-or-Nothing Call&$S_T\mathds{1}_{S_T\geq K}$&$e^{x+\chi}\mathds{1}_{e^{x+\chi-\log K}\geq 1}$\\[10pt]
Asset-or-Nothing Put&$S_T\mathds{1}_{S_T\leq K}$&$e^{x+\chi}\mathds{1}_{e^{x+\chi-\log K}\leq 1}$\\[10pt]
Asymmetric Call&$(S_T^n-K^n)\mathds{1}_{S_T\geq K}$&$K^n(e^{n(x+\chi-\log K)}-1)\mathds{1}_{e^{x+\chi-\log K}\geq 1}$\\[10pt]
Asymmetric Put&$(K^n-S_T^n)\mathds{1}_{S_T\leq K}$&$-K^n(e^{n(x+\chi-\log K)}-1)\mathds{1}_{e^{x+\chi-\log K}\leq 1}$\\[10pt]
Symmetric Call&$(S_T-K)^n\mathds{1}_{S_T\geq K}$&$\sum\limits_{j=0}^n \binom{n}{j} (-1)^{(n-j)}e^{j(x+\chi)+(n-j)\log K}\mathds{1}_{e^{x+\chi-\log K}\geq 1}$\\[10pt]
Symmetric Put&$(K-S_T)^n\mathds{1}_{S_T\leq K}$&$\sum\limits_{j=0}^n \binom{n}{j}(-1)^{(n-j)}e^{(n-j)(x+\chi)+j\log K}\mathds{1}_{e^{x+\chi-\log K}\leq 1}$\\[10pt]
\hline
\end{tabular}} \label{table:CFSPayoff1}
\end{table}

\begin{table}
\small
\caption{Complex Fourier transforms for a variety of financial contingency claims}  
\centering 
\resizebox{\textwidth}{!}{\begin{tabular}{|ccc|} 
\hline
Financial Contingency Claim&Fourier Transform& Fourier Transform\\[10pt]
&$\widehat{G}_k$&$\widehat{G}_0$\\[10pt]
\hline
\rule{0pt}{4ex} 
Call&$\frac{K(d-c)}{i 2\pi k+(d-c)}\left(e^{(i\frac{2\pi}{d-c}k+1)d}-1\right)-\frac{K(d-c)}{i 2\pi k}\left(e^{(i\frac{2\pi}{d-c}k)d}-1\right)$&$K(e^d-1)-K d$\\[10pt]
Put&$\frac{K(d-c)}{i 2\pi k+(d-c)}\left(e^{(i\frac{2\pi}{d-c}k+1)c}-1\right)-\frac{K(d-c)}{i 2\pi k}\left(e^{(i\frac{2\pi}{d-c}k)c}-1\right)$&$K(e^c-1)-K c$\\[10pt]
Covered Call&$-\frac{K(d-c)}{i 2\pi k+(d-c)}\left(e^{(i\frac{2\pi}{d-c}k+1)c}-1\right)+\frac{K(d-c)}{i 2\pi k}\left(e^{(i\frac{2\pi}{d-c}k)c}-1\right)$&$-K(e^c-1)+K c$\\[10pt]
Cash-or-Nothing Call&$\frac{(d-c)}{i 2\pi k}\left(e^{(i\frac{2\pi}{d-c}k)d}-1\right)$&$d$\\[10pt]
Cash-or-Nothing Put&$-\frac{(d-c)}{i 2\pi k}\left(e^{(i\frac{2\pi}{d-c}k)c}-1\right)$&$-c$\\[10pt]
Asset-or-Nothing Call&$\frac{(d-c)}{i 2\pi k+(d-c)}\left(e^{(i\frac{2\pi}{b-a}k+1)d}-1\right)$&$(e^d-1)$\\[10pt]
Asset-or-Nothing Put&$-\frac{(d-c)}{i 2\pi k+(d-c)}\left(e^{(i\frac{2\pi}{d-c}k+1)c}-1\right)$&$-(e^c-1)$\\[10pt]
Asymmetric Call&$\frac{K^n(d-c)}{i 2\pi k+n(d-c)}\left(e^{(i\frac{2\pi}{d-c}k+n)d}-1\right)-\frac{K^n(d-c)}{i 2\pi k}\left(e^{(i\frac{2\pi}{d-c}k)d}-1\right)$&${K^n\over n}(e^{nd}-1)-K^n d$\\[10pt]
Asymmetric Put&$\frac{K^n(d-c)}{i 2\pi k+n(d-c)}\left(e^{(i\frac{2\pi}{d-c}k+1)c}-1\right)-\frac{K^n(d-c)}{i 2\pi k}\left(e^{(i\frac{2\pi}{d-c}k)c}-1\right)$&${K^n\over n}(e^{nc}-1)-K^n c$\\[10pt]
Symmetric Call&$\sum\limits_{j=0}^n \binom{n}{j} (-1)^{(n-j)}\frac{K^n(d-c)}{i 2\pi k+j(d-c)}\left(e^{(i\frac{2\pi}{d-c}k+j)d}-1\right)$&$\sum\limits_{j=1}^n \binom{n}{j} (-1)^{(n-j)}\times$\\[10pt]
&&${K^n\over j}(e^{jd}-1)+(-1)^{n}K^nd$\\[10pt]
Symmetric Put&$\sum\limits_{j=0}^n \binom{n}{j} (-1)^{(n-j+1)}\frac{K^n(d-c)}{i 2\pi k+(n-j)(d-c)}\left(e^{(i\frac{2\pi}{d-c}k+n-j)c}-1\right)$&$\sum\limits_{j=0}^{n-1} \binom{n}{j} (-1)^{(n-j+1)}\times$\\[10pt]
&&${K^n\over n-j}(e^{(n-j)c}-1)-K^nc $\\[10pt]
\hline
\end{tabular}} \label{table:CFSPayoff2}
\end{table}

\section{Singular Fourier-Pad\'e Algorithm and Locating Singularities }\label{sec:SFP_algorithm}
The approach to computing the polynomial coefficients needed in the SFP method is fairly straightforward. To demonstrate the algorithm, we focus on a simple case where the option pricing and Greeks formulae are infinitely smooth apart from jumps located at the endpoints $c$ and $d.$ As we consider $z=\exp\left(i{2\pi\over d-c}y_1\right)$ in either the option pricing formula or the Greeks formula, the jump of $c$ and $d$ in the z-plane is -1. For sake of simplicity, we denote $f_1(z)$ as the CFS representation of any European-style pricing formula or its option Greeks one. With some superscripts dropped for clarity and knowing that $s=1,$ in (\ref{eqn:SFP_1}), we have 
\begin{align}\label{eqn:SFP_al1}
 P_N(z)+ L_{N_1}(z)\log\left(1-{z\over \varepsilon_1} \right)=f_1(z)Q_M(z)+\mathcal{O}(z^{U+1}),
\end{align}
where $N+M+N_1=U.$ Both  $L_{N_1}$ and $f_1(z)$ have Taylor series and CFS expansions, respectively, to determine U; therefore, their expansions are
\begin{align}
\log\left(1-{z\over \varepsilon_s} \right)&=\sum_{k=1}^{U}-{z^k\over \varepsilon_1^k}+0\\
f_1(z)&=2\sum_{k=1}^{U}\widehat{B}_k \widehat{G}_k z^k+ \widehat{B}_0\widehat{G}_0.
\end{align}
Our goal is to derive a linear system for the unknown polynomial coefficients. Note that $Q_M(z)$ and $L_{N_1}(z)$ are determined only by terms of order greater than $N$. Accordingly, we seek a linear solution to 
\begin{align}\label{eqn:SPFMatrix_1}
\begin{bmatrix} 
\widehat{B}\widehat{G} &-L \\ 
\end{bmatrix}
\begin{bmatrix}
\mathbf{q}\\ \mathbf{l}
\end{bmatrix}
=\mathbf{0}.
\end{align}
Here, $\widehat{B}\widehat{G}$ is the $(M+N_1+1)\times(M+1)$ Toeplitz matrix
\begin{align}
\begin{bmatrix} 
\widehat{B}_{\frac{U}{2}+1}\widehat{G}_{\frac{U}{2}+1}&\widehat{B}_{\frac{U}{2}}\widehat{G}_{\frac{U}{2}}& \cdots &\widehat{B}_1\widehat{G}_1 \\ 
\widehat{B}_{\frac{U}{2}+2}\widehat{G}_{\frac{U}{2}+2} &\widehat{B}_{\frac{U}{2}+1}\widehat{G}_{\frac{U}{2}+1}  & \ddots & \widehat{B}_2\widehat{G}_2\\ 
\vdots&\vdots &\ddots &\vdots\\
\widehat{B}_U\widehat{G}_U & \widehat{B}_{U-1}\widehat{G}_{U-1} &\cdots & \widehat{B}_{\frac{U}{2}}\widehat{G}_{\frac{U}{2}},
\end{bmatrix}
\end{align}
and L is the $(M + N_1 + 1) � (N_1 + 1)$ matrix defined similarly using the Taylor coefficients of log(1+z).
The vectors $\mathbf{q}=\{q_m\}_{m=0}^{M}$ and $\mathbf{l}=\{l_{n_1}\}_{{n_1}=0}^{N_1}$ hold the unknown polynomial coefficients in order of increasing degree.
As the column dimension of the matrix in (\ref{eqn:SPFMatrix_1}) is one greater than its row dimension, we can conclude that there is one nonzero solution to (\ref{eqn:SPFMatrix_1}). In many cases,  this can be made into a square system by choosing, say, $q_0 = 1$. However, if one does not want to assume that any particular coefficient is nonzero, one can solve (\ref{eqn:SPFMatrix_1}) by a singular value decomposition \citep[cf.][]{Gon_Tre:2013}. Finally, the unknown coefficients of $\mathbf{p}=\{p_n\}_{n=1}^N$ can be obtained by multiplication through the following matrix system:
\begin{align}
\mathbf{p}
=\begin{bmatrix} 
\widehat{B}_0\widehat{G}_0 &  &   &   \\ 
\widehat{B}_1\widehat{G}_1  & \widehat{B}_0\widehat{G}_0 &  & \\ 
\vdots&\ddots &\ddots &\\
\widehat{B}_{{U\over 2}}\widehat{G}_{{U\over 2}}  & \cdots &\cdots& \widehat{B}_0\widehat{G}_0 
\end{bmatrix}
\mathbf{q}
-
\begin{bmatrix} 
l_0 &  &   &   \\ 
l_1 & l_0 &  & \\ 
\vdots&\ddots &\ddots &\\
l_{{U\over 2}}  & \cdots &\cdots&\ l_0
\end{bmatrix}
\mathbf{l}.
\end{align}
If there is more than one jump location in the option pricing/Greeks curve (\ref{eqn:SFP_al1}), this suggests the following modification of the equation: 

\begin{align}
 P_N(z)+ L_{N_1}(z)\log\left(1-{z\over \varepsilon_1} \right)+\ldots+ L_{N_s}(z)\log\left(1-{z\over \varepsilon_S} \right)=f_1(z)Q_M(z)+\mathcal{O}(z^{U+1}). 
\end{align}
Accordingly, we have to modify (\ref{eqn:SPFMatrix_1}) to produce a new $L$ matrix and a vector of coefficients for each location to reflect the changes. According to \cite{Dris_Ben:2001, Dris_Ben:2011}, there is no rigorous optimal formula for choosing the degrees $M,$ $N,$ and $N_1,$\ldots,$N_s.$ Because the denominator polynomial $Q_M$ is shared, we allow $M$ to be the largest, with the others being equal as far as possible. For the case of just one jump location, taking $N$ at roughly $40\%$ of the total available degrees of freedom seems to work well. Experiments suggest that these choices can affect the observed accuracy, occasionally by as much as an order of magnitude, but on average, there is little variation within a broad range of choices. 

The methods described above all assume that the locations of all jumps are known in the option pricing/Greeks curve. For jumps whose locations are unknown, we follow the suggestion of \citet{Dris_Ben:2001, Dris_Ben:2011} to use the Fourier-Pad\'e algorithm to estimate their locations. As we approximate a PDF rather than a payoff function with the CFS method (see the proof of Theorem \ref{EuroCallOption}), we only consider jumps existing in the PDF. The existence of a jump is attributed to a combination parameters and/or a very short maturity that causes a sharp-peaked PDF curve. 

Using the Fourier-Pad\'e algorithm (cf. the Fourier-Pad\'e approximant in Section \ref{sec:SFP}) to estimate the jumps in a PDF is fairly simple. We first express the PDF as the CFS representation:  
\begin{align}\label{eqn:CFS_PDF_temp2}
\mathfrak{Re}\left[2\sum_{k=1}^{\infty} \varphi\left(-\frac{2\pi}{d-c}k\right) e^{i\frac{2\pi}{b-a}ky}+\varphi\left(0\right)\right].
\end{align}
Then, we can differentiate (\ref{eqn:CFS_PDF_temp2}) with respect to $y$ to obtain \begin{align}
\mathfrak{Re}\left[2\sum_{k=1}^{\infty} \left(i\frac{2\pi}{b-a}k\right) \varphi\left(-\frac{2\pi}{d-c}k\right) e^{i\frac{2\pi}{b-a}ky}\right].   
\end{align}
Finally, we let $z=\exp\left({i\frac{2\pi}{d-c}y}\right)$ in the two equations above, and they are ready for the Fourier-Pad\'e approximation. In general, when the PDF has a jump, the sharp-peaked jump point will have an enormously large value after differentiation. Then, based on this result, we find that point in the original PDF. In other words, Figure \ref{fig:VG_SmRou_SFP} is a graphical illustration of the outlooks of two PDFs (top) under the VG model (cf. Section \ref{subsection:VG}) and the first derivative (bottom) after the Fourier-Pad\'e approximation. In the figure, we can see that the PDFs in the top and bottom panels are smooth without any jumps, as they do not show any point with an extremely large value. However, in the graph on the bottom right, we can see that there is a jump that produces a value of $15\times 10^4.$ 
\begin{figure}
\center
\includegraphics[height=6cm,width=10cm]{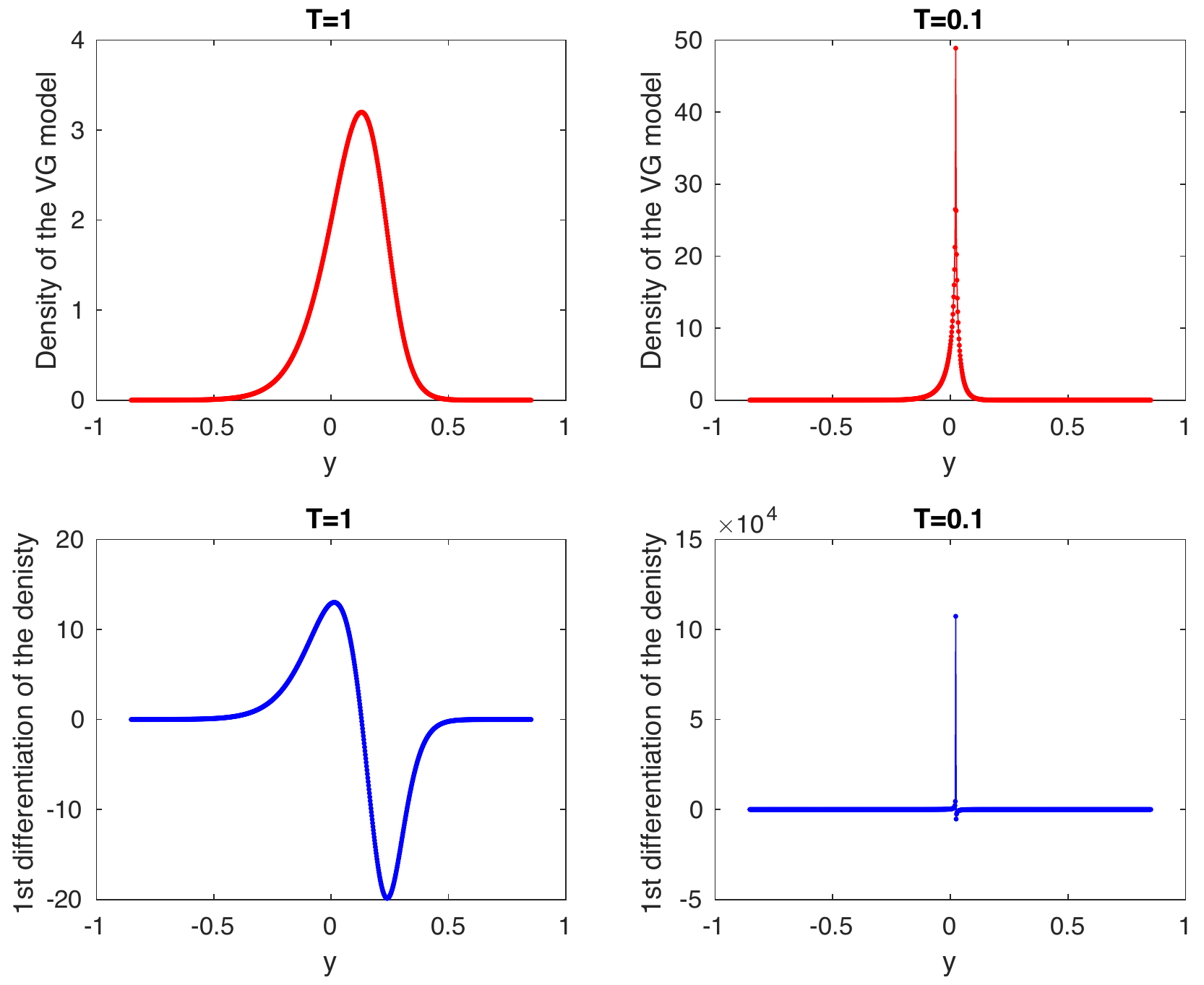}
\setlength{\abovecaptionskip}{1pt}
\caption{Density functions (top) of the VG model and their first derivative (bottom). A piecewise continuous (non-smooth) PDF with a jump after/before differentiation is shown in the graphs on the top and bottom right. The parameters are taken from \textbf{VG--Para1}.}
\label{fig:VG_SmRou_SFP}
\end{figure}


\section{Choice of Truncated Intervals}\label{sec:trunc}
As we will show in Section \ref{sec:error}, the choice of the interval $[c,d]$ is affected by the accuracy of the SFP method. A minimum and substantial interval $[c,d]$ can capture most of the mass of a PDF such that the SFP method can in turns produce sensible global spectral convergence rate. In this short section, we show how to construct an interval related to the closed-form formulas of stochastic process cumulants. The idea of using the cumulants is first proposed by \citet{Fan_Oos:2009a} to construct the definite interval $[c,d]$ in (\ref{eqn:charAppx}). Based on their ideas, we have the following expression for $[c,d]$:
\begin{align}\label{eqn:truncInt1}
d&= \left\vert c_1+L\sqrt{c_2+\sqrt{c_4}}+\left\vert \log\left({S_0\over K}\right)\right\vert \right\vert\\
c&=-d,
\end{align}
where $c_1,$ $c_2,$ and $c_4$ are the first, second and fourth cumulants, respectively, of the stochastic process and $L\in[10,12].$ For simple, less-complicated financial models, we also obtain closed-form formulas for $c_1,$ $c_2,$ and $c_4$, which are shown in Table \ref{table:cumulants} of Appendix \ref{section:cums}. However, in the Heston model, we use the absolute value of $c_2$ and ignore the value of $c_4$ due to the negative value of $c_2$ and the lengthy representation of $c_4$ \citep[cf.][]{Fan_Oos:2009a}. We therefore have $c_1+L\sqrt{|c_2|}$ rather than $c_1+L\sqrt{c_2+\sqrt{c_4}}.$  The truncated intervals only work for smooth PDFs without any jumps. In general, by trial and error, if there is a jump in a PDF, we add 0.5 to (\ref{eqn:truncInt1}) to allow the better convergence of our method. Accordingly, the formulae can be transformed as: 
\begin{align}\label{eqn:truncInt2}
d&= \left\vert c_1+L\sqrt{c_2+\sqrt{c_4}}+\left\vert \log\left({S_0\over K}\right)\right\vert\right\vert+0.5\\
c&=-d.
\end{align}
\begin{remark}\label{remarks:intervals}
Without a doubt, in most cases, the truncated interval of (\ref{eqn:truncInt2}) works for any normal non-smooth PDF. However, when the conditions become very extreme, for example, $T=1e-06$ in \textbf{BSM--Para4} in Section \ref{sec:BSM}, the PDF is extremely narrow, thin and spiky at one point. The interval of (\ref{eqn:truncInt2}) can still perform but is not optimal. To improve it, we replace $0.5$ in the equation with $0.1.$ This is attributed to the desire for higher accuracy and convergence with fewer terms required over a short interval range. 
\end{remark}
\section{Error Analysis}\label{sec:error}
In this section, we demonstrate that the total error from pricing European-style options can be made very small by choosing a suitably large interval $[c,d].$ Furthermore, we show that global spectral convergence can be achieved at any point along the option pricing curve even though the input PDF is a $\mathbf{C}^\nu$ piecewise continuous (non-smooth) function.

In this paper, there are three types of approximation errors in any call/put option.
\begin{enumerate}
\item Integration truncation error:
\begin{align}\label{error1}
\epsilon_1:=\left\vert\int_{-\infty}^\infty  G(e^{x+\chi}, K)f(\chi)  \mathrm{d} \chi -\int_{c}^d G(e^{x+\chi}, K) f(\chi) \mathrm{d} \chi \right\vert
\end{align}
\item 
Error related to approximating (\ref{eqn:EuroCall_CFS_temp}) with (\ref{eqn:EuroCall_CFS}):
\begin{align}\label{error2}
\epsilon_2&:=\Bigg\vert \mathfrak{Re}\left[2\sum_{k=1}^{\infty}\frac{1}{d-c}\int_{c}^{d} f(y)e^{-i\frac{ 2\pi }{d-c}ky} \mathrm{d}y\,\widehat{G}_k\,e^{i\frac{2\pi}{d-c}k\left(-x+\log K\right)} + \frac{1}{d-c}\int_{c}^{d} f(y)\mathrm{d}y\,\widehat{G}_0\right]-\nonumber \\
&\quad\quad\mathfrak{Re}\left[2\sum_{k=1}^{\infty}\widehat{B}_k\,\widehat{G}_k\,e^{i\frac{2\pi}{d-c}k\left(-x+\log K\right)} + \widehat{B}_0\,\widehat{G}_0\right]\Bigg\vert\nonumber\\
&=\Bigg\vert \mathfrak{Re}\Bigg[2\sum_{k=1}^{\infty}\left(\frac{1}{d-c}\int_{c}^{d} f(y)e^{-i\frac{ 2\pi }{d-c}ky} \mathrm{d}y-\widehat{B}_k\right)\,\widehat{G}_k\,e^{i\frac{2\pi}{d-c}k\left(-x+\log K\right)} + \nonumber\\
&\quad\quad\left(\frac{1}{d-c}\int_{c}^{d} f(y)\mathrm{d}y-\widehat{B}_0\right)\,\widehat{G}_0\Bigg]\Bigg\vert
\end{align}
\item Truncated SFP series error:
\begin{align}\label{error3}
\epsilon_3&:=\left\vert\mathfrak{Re}\left[2\sum_{k=1}^{\infty} \widehat{B}_k\widehat{G}_k z^k+\widehat{B}_0\widehat{G}_0\right] -\mathfrak{Re}\left[{ P_N^+(z)+\sum_{s=1}^S L^+_{N_s}(z)\log\left(1-z/\varepsilon_s \right) \over Q^+_M(z)} \right]\right\vert,\nonumber\\
&z=e^{i{2\pi\over d-c}(-x+\log K)} \hbox{ and } \varepsilon_s=e^{i{2\pi\over d-c}(\zeta_s)} 
\end{align}
\end{enumerate}

If we introduce the concept of the cumulative probability density function (CDF) $F(\chi)$ such that $f(\chi)d\chi=d F(\chi),$ we can simplify the integration truncation error as follows:
\begin{align}
\epsilon_1&=\left\vert\left(\int_{-\infty}^\infty  G(e^{x+\chi},K) f(\chi) \mathrm{d} \chi -\int_{c}^d G(e^{x+\chi},K) f(\chi) \mathrm{d} \chi\right)\right\vert\nonumber\\
&=\left\vert\left(\int_{-\infty}^c  G(e^{x+\chi},K) f(\chi) \mathrm{d} \chi +\int_{d}^\infty G(e^{x+\chi}) f(\chi) \mathrm{d} \chi\right)\right\vert\nonumber\\
&\leq\left\vert\left(\int^{c}_{-\infty}  {\partial G(e^{x+\chi},K)\over \partial \chi} F(\chi) \mathrm{d} \chi\right) \right\vert + \left\vert \int_{d}^\infty {\partial G(e^{x+\chi},K) \over \partial \chi} (1-F(\chi)) \mathrm{d} \chi \right\vert\\
&\approx 0: \quad(\hbox{if}\, \chi=c, d,-\infty, \infty).
\end{align}
We can see that $\epsilon_1$ is bounded and approaches zero as long as $[c,d]$ is chosen reasonably such that $1-F(d)\approx 0$ when $d< \infty $ or $F(c)\approx 0$ when $c> -\infty.$ We are also able to adapt the same idea to investigate the bound of $\epsilon_2.$ Accordingly, taking into account $\vert \exp(i{2\pi k\over d-c}y)\vert\leq 1,$ we first investigate the error
$$\underline{\epsilon}_2:=\left\vert {1 \over d-c}\int_{c}^{d} f(y)e^{-i\frac{ 2\pi }{d-c}ky} \mathrm{d}y-\widehat{B}_k\right\vert$$ 
in $\epsilon_2.$ If we expand the equation above, we obtain 
\begin{align}
\underline{\epsilon}_2&:=\left\vert {1 \over d-c}\int_{c}^{d} f(y)e^{-i\frac{ 2\pi }{d-c}ky} \mathrm{d}y- {1 \over d-c}\varphi\left(-i\frac{ 2\pi }{d-c}k\right)\right\vert\\
&=\left\vert {1 \over d-c}\int_{-\infty}^{\infty} f(y)e^{-i\frac{ 2\pi }{d-c}ky} \mathrm{d}y- {1 \over d-c}\int_{c}^{d} f(y)e^{-i\frac{ 2\pi }{d-c}ky} \mathrm{d}y\right\vert\\
&\leq\left\vert {1 \over d-c}\left(\int_{-\infty}^{c} f(y)\mathrm{d}y+\int_{d}^{\infty} f(y)\mathrm{d}y\right)\right\vert\\
&=\left\vert{1 \over d-c}\left(F(\infty)-F(d)+F(c)-F(-\infty)\right)\right\vert\\
&\approx 0: \quad(\hbox{if}\, y=c, d,-\infty, \infty).
\end{align}
Based on the result above,
\begin{align} 
\epsilon_2&:=\Bigg\vert \mathfrak{Re}\Bigg[2\sum_{k=1}^{\infty}\left(\frac{1}{d-c}\int_{c}^{d} f(y)e^{-i\frac{ 2\pi }{d-c}ky} \mathrm{d}y-\widehat{B}_k\right)\,\widehat{G}_k\,e^{i\frac{2\pi}{d-c}k\left(-x+\log K\right)} + \nonumber\\
&\quad\quad\left(\frac{1}{d-c}\int_{c}^{d} f(y)\mathrm{d}y-\widehat{B}_0\right)\,\widehat{G}_0\Bigg]\Bigg\vert\nonumber\\
&\leq \Bigg\vert \mathfrak{Re}\Bigg[2\sum_{k=1}^{\infty} \underline{\epsilon}_2 \,\widehat{G}_k\,e^{i\frac{2\pi}{d-c}k\left(-x+\log K\right)} + \left(\frac{1}{d-c}\int_{c}^{d} f(y)\mathrm{d}y-\frac{1}{d-c}\int_{c}^{d} f(y)\mathrm{d}y \right)\,\widehat{G}_0\Bigg]\Bigg\vert\nonumber\\
&\approx 0.
\end{align}
To conclude that $\epsilon_2$ is approaching zero, we first note that there is no approximate error of $G_k$ and $G_0$ because of their closed-form expressions. Then, once $\underline{\epsilon}_2$ tends to zero, the first term of the equation will also diminish to zero. The last term of the equation tends to zero because  $\widehat{B}_0$ equals $\frac{1}{d-c}\int_{c}^{d} f(y)\mathrm{d}y$ (cf. (\ref{eqn:B_k}) and (\ref{eqn:BkGk})).

Finally, the SPF series truncation error is also bounded \citep[cf.][]{Dris_Ben:2001, Dris_Ben:2011} and can be formulated as follows:
\begin{align}
\epsilon_3&:=\left\vert\mathfrak{Re}\left[2\sum_{k=1}^{\infty} \widehat{B}_k\widehat{G}_k e^{i\frac{2\pi}{d-c}k(-x+\log K)}+\widehat{B}_0\widehat{G}_0\right] -\mathfrak{Re}\Bigg[{ P_N^+(z)+\sum_{s=1}^S L^+_{N_s}(z)\log\left(1-z/\varepsilon_s \right) \over Q^+_M(z)} \Bigg]\right\vert\nonumber\\
&=\mathcal{O}(z^{N+M+\sum_{s=1}^{N_s} N_s+1}).
\end{align}
The error term of $\epsilon_3$ tends to zero with a global spectral rate of $\mathcal{O}(z^{N+M+\sum_{s=1}^{N_s} N_s+1})$ even though the input PDF is $C^{\nu}$ piecewise continuous. 

Before we illustrate the total error bound when approximating any true European-type option price $V(x,K,t)$ defined as 
\begin{align}
\epsilon&:=\left\vert V(x,K,t)-e^{-r(T-t)}\mathfrak{Re}\left[ P_N^+(z)+\sum_{s=1}^S L^+_{N_s}(z)\log\left(1-z/\varepsilon_s \right) \over Q^+_M(z)\right]\right\vert,
\end{align}
we first summarise the whole approximation procedure of European-type option prices and note where $\epsilon_1,$ $\epsilon_2,$ and $\epsilon_3$ lie. We start off by seeking a definite interval $[c,d]$ that allows us to approximate $V(x,K,t)$ defined on $[-\infty,\infty]$ in (\ref{eqn:GEquation_1}) with the form
$$V(x,K, t)\approx e^{-r(T-t)}\int_{c}^{d} G(e^{x + \chi},K) f(\chi) \mathrm{d}\chi.$$
The interval $[c,d]$ we proposed satisfies condition (\ref{eqn:charAppx}). As a result, we obtain our first approximation error $\epsilon_1.$ As $V(x,K,t)$ is now approximated in $[c,d],$ this implies that we can construct a CFS expansion of $V(x,K,t),$ like the one in (\ref{eqn:EuroCall_CFS_temp}). Then, because including a characteristic function $\varphi(\cdot)$ in the CFS expansion allows for a more accurate approximation, we have another CFS expansion of $V(x,K,t),$ given as in (\ref{eqn:EuroCall_CFS}). Accordingly, we have $\epsilon_2,$ an approximation error of  (\ref{eqn:EuroCall_CFS_temp}), being approximated by (\ref{eqn:EuroCall_CFS}). Finally, $\epsilon_3$ is the error of (\ref{eqn:EuroCall_SPF1}), which is the formula for approximating (\ref{eqn:EuroCall_CFS}) with the SFP approximant (\ref{eqn:SFP_1}).

By combining the results of $\epsilon_1,$ $\epsilon_2$ and $\epsilon_3,$ we can determine the total error bound $\epsilon$; hence, we have an inequality of  
\begin{align}\label{totalerror1}
\epsilon&=\left\vert V(x,K,t)-e^{-r(T-t)}\mathfrak{Re}\left[ P_N^+(z)+\sum_{s=1}^S L^+_{N_s}(z)\log\left(1-z/\varepsilon_s \right) \over Q^+_M(z)\right]\right\vert\\
&=\left\vert e^{-r(T-t)}\Bigg(\int_{-\infty}^{\infty}G(e^{x+\chi},K) f(\chi) \mathrm{d} \chi- \mathfrak{Re}\left[ P_N^+(z)+\sum_{s=1}^S L^+_{N_s}(z)\log\left(1-z/\varepsilon_s \right) \over Q^+_M(z)\right]\Bigg)\right\vert\nonumber\\
&\leq \bigg\vert e^{-r(T-t)}\bigg\vert\Bigg(\left\vert\int_{-\infty}^\infty G(e^{x+\chi},K) f(\chi) \mathrm{d} \chi -\int_{c}^d G(e^{x+\chi}, K) f(\chi)\mathrm{d} \chi\right\vert+\nonumber\\
&\quad \Bigg\vert \mathfrak{Re}\left[2\sum_{k=1}^{\infty}\frac{1}{d-c}\int_{c}^{d} f(y)e^{-i\frac{ 2\pi }{d-c}ky} \mathrm{d}y\,\widehat{G}_k\,e^{i\frac{2\pi}{d-c}k\left(-x+\log K\right)} + \frac{1}{d-c}\int_{c}^{d} f(y)\mathrm{d}y\,\widehat{G}_0\right]-\nonumber\\
&\quad\mathfrak{Re}\left[2\sum_{k=1}^{\infty} \widehat{B}_k\widehat{G}_k e^{i\frac{2\pi}{d-c}k(-x+\log K)}+\widehat{B}_0\widehat{G}_0\right]\Bigg\vert \nonumber+\\
&\quad \left\vert \mathfrak{Re}\left[2\sum_{k=1}^{\infty} \widehat{B}_k\widehat{G}_k z^k+\widehat{B}_0\widehat{G}_0\right]- \mathfrak{Re}\left[ P_N^+(z)+\sum_{s=1}^S L^+_{N_s}(z)\log\left(1-z/\varepsilon_s \right) \over Q^+_M(z) \right]\right\vert\Bigg) \nonumber\\
&\leq \vert e^{-r(T-t)}\vert(\epsilon_1+\epsilon_2+\epsilon_3)\vert\nonumber\\
&< \vert e^{-r(T-t)}\vert(\epsilon_1+\epsilon_2+\mathcal{O}(z^{N+M+\sum^S_{s=1} N_s+1}))\vert\nonumber\\
&\approx 0.
\end{align}
\begin{remark}
According to \cite{Dris_Ben:2001, Dris_Ben:2011}, the rate of $\mathcal{O}(z^{N+M+\sum_{s=1}^{N_s} N_s+1})$ is not spatially uniform because convergence at a jump is somewhat limited by the well-known numerical ill-conditioning of the straightforward Pad\'e problem. Moreover, when a curve has a discontinuous point (zeroth-order jump), it also has a convergence rate that is the average of the one-sided limit values. Through the numerical experiments in \cite{Dris_Ben:2001}, we can see that the SFP method can yield spectral convergence rate at jumps. This is not necessarily the case if the jumps are very difficult to interpolate. Nevertheless, as \cite{Dris_Ben:2011} suggest, we can still have 4--6 digits of accuracy at the jumps when they are known in advance. When the jumps are not known in advance, we can use the Fourier-Pad\'e algorithm to find them (see Section \ref{sec:SFP_algorithm}). By incorporating this technique, the SFP method still can perform and yield spectral convergence or 4th-order convergence  when the jumps are very difficult to interpolate. 
\end{remark}

\section{Numerical Results}\label{sec:results}
In this section, we demonstrate the performance of the SFP method through various numerical tests. The purpose of this section is first to test whether the error convergence analysis presented in Section \ref{sec:error} is in line with the numerical findings in this section. Second, we test the ability of the SFP method to price any European-style option that are deep in/out of the money and have long/short maturities. Third, we analyse whether the SFP method can provide consistent accuracy when approximating small or large values of option prices. Finally, in a major development of the method, we test whether the SFP method can retain global spectral convergence even when the PDF is piecewise continuous. A number of popular numerical methods are implemented to test the SFP method in terms of the error convergence, convergence rate and computational time. These methods include the COS method (a Fourier COS series method, \citealp{Fan_Oos:2009a}), the filter-COS method (a COS method with an exponential filter to resolve the Gibbs phenomenon; see \citealp{Ruj_Oos:2013}), the CONV method (an FFT method, \citealp{Lor_Fan:2008}), the Lewis-FRFT method (a fractional FFT method, \citealp{Lew:2001, Cho:2004}), and the B-spline, Haar wavelet and the SWIFT methods (a wavelet-based method; see \citealp{Gra_Oos:2013, Gra_Oos:2016}). When we implement the CONV and Lewis-FRFT methods, we use Simpson's rule for the Fourier integrals to achieve fourth-order accuracy. In the filter-COS method, we use an exponential filter and set the accuracy parameter to $10$ as \cite{Ruj_Oos:2013} report that this filter provides better algebraic convergence than the other options. We also set the damping factors of the CONV and Lewis-FRFT to 0 and any value greater than zero, respectively.

As the SFP method requests approximating jumps in logarithmic series, we consider and apply the endpoints $c$ and $d$ as our two known jumps for all non-smooth/smooth PDFs. Only the jump of the non-smooth PDF of the Black-Scholes-Merton (BSM) model is known as its mean value. For the rest of the non-smooth PDFs, we use the Fourier-Pad\'e algorithm (cf. Section \ref{sec:SFP_algorithm}) to find their locations. In all numerical experiments, we use the parameter $U$ to denote the number of terms of the SFP method and $N$ to denote the number of terms/grid points of the others. When we measure the approximation errors of the numerical methods, we use absolute errors, the infinity norm errors $R_\infty$ and the $L_2$ norm errors $R_2$ as the measurement units. Moreover, to improve the accuracy of our method, we use the call-put parity--$V_{call}(x,K,t) = V_{put}(x,K,t) + S_0\exp(-qT)-K\exp(-rT)$--to approximate call prices once we have put prices ready. Finally, all the CPU times presented (in seconds) are determined after averaging the computational time over 120 experiments. A MacBook Pro with a 2.8 GHz Intel Core i7 CPU and two 8 GB DDR SDRAM (cache memory) is used for all experiments. The code is written in MATLAB R2011b. Finally, the MATLAB code of implementing the COS method and the FFT method, such as the CONV method and the like, is retrieved from \citet{BEN:2015}.

\subsection{Exponential L\'evy Processes}
\subsubsection{The Black-Scholes-Merton Model}\label{sec:BSM}
The first numerical experiments are performed using the BSM model \citep[cf.][]{Bla_Sch:1973, Mer:1973}. The stock dynamics driven by the BSM model (a geometric Brownian process) are given by 
\begin{eqnarray}\label{eqn:Brownian}
S_T=S_0e^{(r-q-{1\over 2}\sigma^2)T+\sigma W_T},
\end{eqnarray}
where $W_{T}$ is a risk-neutral Brownian motion, and $\sigma$ is the volatility. The characteristic function of the model is also defined as
\begin{eqnarray}\label{eqn:Brownian_charexponent}
\varphi(u)=\exp\left(T\left(iu(r-q-{1\over 2}\sigma^2)-\frac{1}{2}\sigma^2u^2\right)\right), \,\, u\in\mathbb{R}.
\end{eqnarray}
The parameters for the experiments are selected from the following:
\begin{align}
\textbf{BSM--Para1}:\,S_0&= 100,\, \sigma=0.15,\, r= 0.03,\, T = 1.0,\, q= 0.0,\\ 
\textbf{BSM--Para2}:\,S_0&=100,\, \sigma=0.25,\, r=0.1,\,T= 50 \mbox{ or } 100,\, q=0, \, K=120,\\  
\textbf{BSM--Para3}:\, K &=100,\, \sigma=0.2,\, r=0.06,\, T=1e-06 ,\, q=0.
\end{align}
In the first numerical test (\textbf{BSM--Para1}), we first check for convergence behaviour against a range of strikes $K$ from $1$ to $200$ for deep in/out-of-the money and at-the-money vanilla put options, respectively. The parameters are retrieved from \cite{BEN:2015}. Using the same parameters, we check for convergence behaviour against a range of strikes $K$ from $80$ to $120$ for cash-or-nothing put options rather than vanilla options in our second numerical test. We compare our method with the COS method, the CONV method and the Lewis-FRFT method in the first test but only with the COS method in the second test. The third numerical experiment (\textbf{BSM--Para2}) is devoted to comparing the performance of the COS method, the SFP method, the SWIFT method for long maturity call options. As we sometimes encounter these options in the insurance and pension industry, it is worth testing our method against them. The parameters are retrieved from \cite{Gra_Oos:2016}} for the test. Finally, for the last numerical test (\textbf{BSM--Para3}), we use the values of $\sigma,$ $r$ and $q$ from \cite{And_Wid:2003} and check for convergence behaviour against a range of stock prices $S_0$ from $80$ to $120$ for very short maturity vanilla call options and their option Delta and Gamma. Our method is compared with the COS, filter-COS, Lewis-FRFT and CONV methods. Reference values for all the tests are based on the BSM analytical formula. In each numerical test, except for the third one, we declare 250 different option prices within the range of either $K$ or $S_0$ to test the efficiency of our method and the others. We represent all recovered PDFs via the SFP method in Figure \ref{fig:BSMDensitySFP} for all sets of parameters. We can see that the non-smooth recovered PDF (top right) can be obtained via \textbf{BSM--Para3}. We set $L=10$ in the truncated interval proposed in Section \ref{sec:trunc}. One should note that for the truncated interval in the last numerical test \textbf{BSM--Para3}, we replace $0.5$ with $0.1$ in (\ref{eqn:truncInt2}) (see Remark \ref{remarks:intervals}), as we want our method to have higher accuracy and a better convergence rate. It is quite rare for the maturity time to be infinitesimal as in  \textbf{BSM--Para3}, but to demonstrate the efficiency of our method, it is worth performing such a test.

All the tables in this section and the others suggest that the difference in the computational time across methods is not large. It takes less than 0.1 seconds to approximate $250$ option prices for any method when $N$ and $U$ equal 64. Without considering the accuracy of the methods, this is a quite reasonable time frame in which to produce a substantial number of option prices at once. However, when we consider the error convergence and convergence rate of the methods, there are sizeable differences among them. Figure \ref{fig:BSLRPara1} and its numerical presentation in Table \ref{table:BS_DOM_SFP} are dedicated to the first numerical test (\textbf{BSM--Para1}). In this test, the SFP method has the smallest $R_\infty$ and $R_2$ errors, with $R_\infty$ and $R_2$ less than $5.8e-13$ as $U=64$ when the strike increases from 1 to 200. It also yields global spectral convergence against which none of the other methods can compete. This suggests that the truncated range proposed in this paper can work over longer range than that proposed in \citet{Fan_Oos:2009a}. In Table \ref{table:BS_AllorNo_SFP}, we use the same parameters (\textbf{BSM--Para1}) to examine the ability of the COS and SFP methods to price cash-or-nothing put options. Again, the two methods can achieve very low convergence error and spectral convergence. Table \ref{table:BS_LongMa_SFP} refers to the third test (\textbf{BSM--Para2}) and replicates Table 3 in \cite{Gra_Oos:2016}. In this test, the SFP method impressively provides high accuracy over the SWIFT and COS methods with fewer terms required. For example, when $T=50,$ we only need 32 terms to obtain the true solution with 7-digit accuracy in the SFP method, whereas the other two require more terms to obtain 1-digit accuracy. Finally, in the last test (\textbf{BSM--Para4}), as the cumulants are too small to create a meaningful truncated range for the COS and filter-COS methods, we use (\ref{eqn:truncInt2}) to allow the method to generate relevant option prices. In Figure \ref{fig:BS_shortT_SFP}, the SFP method dominates the other four methods to yield the global spectral convergence rate away from the jump, while the input non-smooth PDF (see top right, Figure \ref{fig:BSMDensitySFP}) behaves like a Dirac Delta function\footnote{The Dirac Delta function is a generalised function or distribution on the real number line that is zero everywhere except at zero.} and has an almost infinitely thin spike near the origin. In Table \ref{table:BS_Sing_SFP_individuals}, when we test the accuracy of the option prices around/at the jump---$S_0=99.999$---the SFP method dominate the other methods. It has zero absolute error in the case of $ S_0=95$ and $6.269e-05$ error from the true solution in the case of $S_0=99.999$ when $U=64.$ This is in line with the findings of \cite{Dris_Ben:2011} as the SFP offers 4-6 digit accuracy at the jumps when the function is very difficult. Obviously, the non-smooth PDF in the last test is very difficult to approximate, as it behaves like a Dirac Delta. Finally, using the same parameters of \textbf{BSM--Para3}, we recover the call Delta and Gamma via the SFP approximant in Figure \ref{fig:BSDeltaGammaSFP}. In the graphs, the SFP provides a solution of extremely high accuracy and global spectral convergence  apart from the jump.

\begin{figure}
\center
\includegraphics[height=6cm,width=10cm]{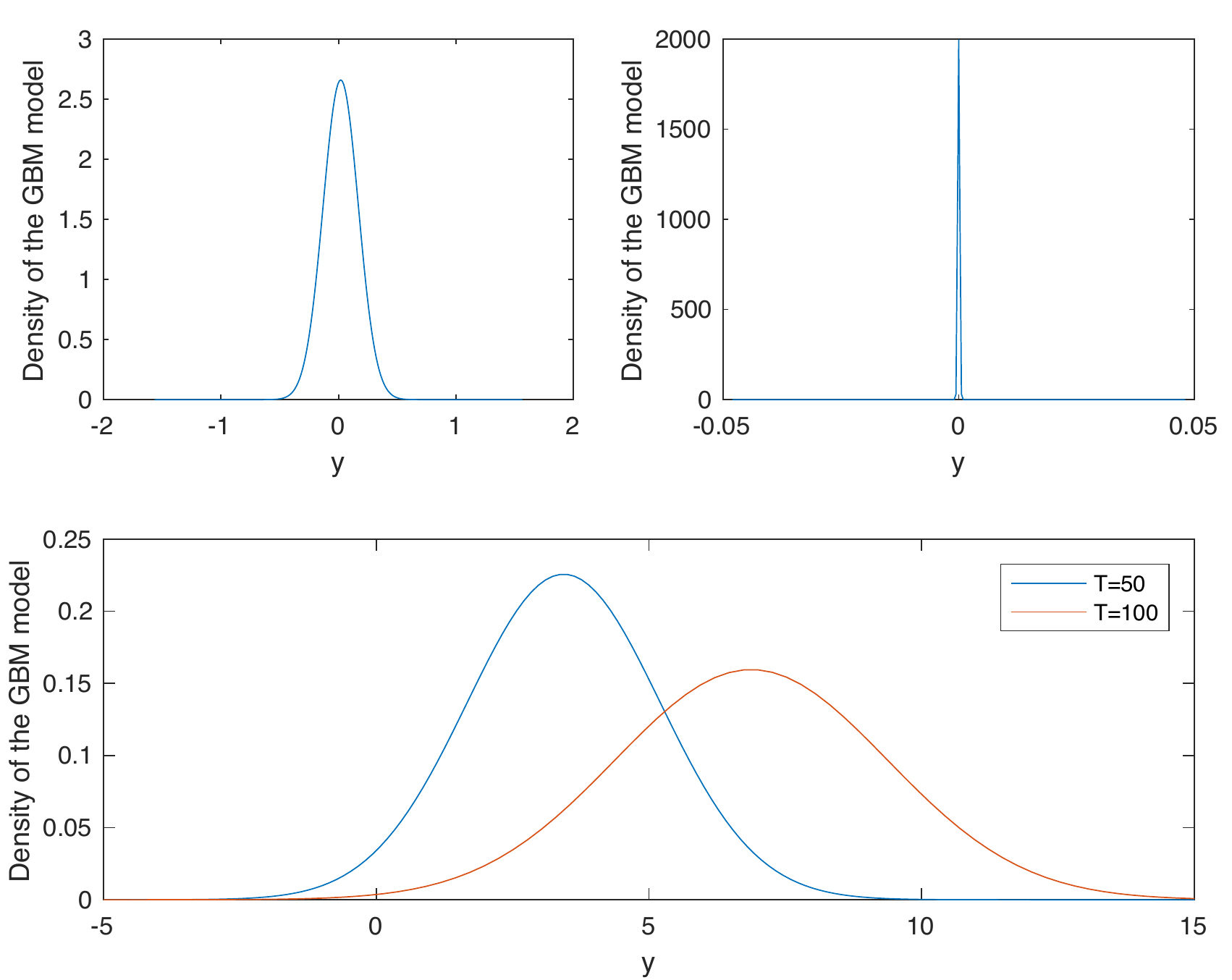}
\setlength{\abovecaptionskip}{1pt}
\caption{Probability density functions are generated by three sets of parameters: \textbf{BSM--Para1} (top left), \textbf{BSM--Para3} (top right) and \textbf{BSM--Para2} (bottom).}
\label{fig:BSMDensitySFP}
\end{figure}

\begin{figure}
\center
\includegraphics[height=6cm,width=10cm]{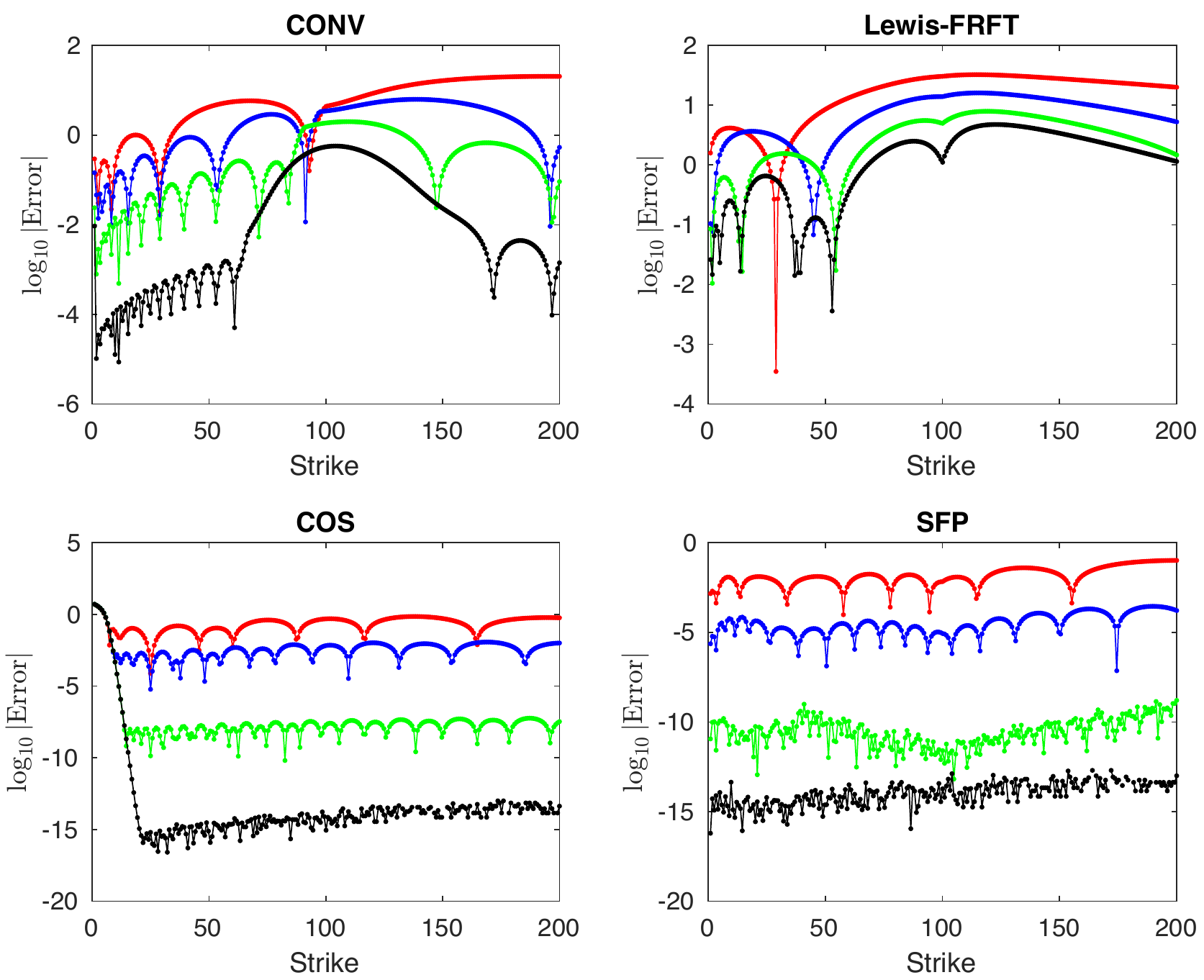}
\setlength{\abovecaptionskip}{1pt}
\caption{Error envelopes of the CONV, Lewis-FRFT, COS and SFP methods for pricing European deep at/in/out-of-the-money put options under the BSM model with strikes increasing from 1 to 200. The number of terms/grid points increases in a sequence of  8 (red line), 16 (blue line), 32 (green line) and 64 (black line). Spectral convergence is observed in the SFP method. The parameters are taken from \textbf{BSM--Para1}.}
\label{fig:BSLRPara1}
\end{figure}

%

\begin{table} 
\caption{Comparison of the CONV, Lewis-FRFT, COS and SFP methods in error convergence and CPU time for pricing European deep at/in/out-of-the-money put options under the BSM model. The range of $K$ is from 1 to 200. Spectral convergence is observed in the SFP method. The parameters are taken from \textbf{BSM--Para1}.} 
\label{table:BS_DOM_SFP}
\centering 
\begin{tabular}{|c|cc|c||c|cc|c|} 
\hline
\multicolumn{4}{|c||}{\textbf{CONV}}&\multicolumn{4}{c|}{\textbf{Lewis-FRFT}}\\
\hline
N&$R_\infty$&$R_2$&Time&N&$R_\infty$&$R_2$&Time\\
\hline
8&20.49&180.1&0.00806&8&32.53&356&0.02111\\
16&6.292&53.54&0.00941&16&16.1&156.4&0.02525\\
32&1.998&13.83&0.01012&32&7.933&69.38&0.04571\\
64&0.5709&3.286&0.01641&64&4.765&38.81&0.09151\\
\hline
\multicolumn{8}{c}{\textbf{}}\\
\hline
\multicolumn{4}{|c||}{\textbf{COS}}&\multicolumn{4}{c|}{\textbf{SFP}}\\
\hline
N&$R_\infty$&$R_2$&Time&U&$R_\infty$&$R_2$&Time\\
\hline
8&5.046&10.06&0.00706&8&1.028e-01&5.893e-01&0.00901\\
16&5.109&8.447&0.00841&16&2.818e-04&1.559e-03&0.00951\\
32&5.111&8.449&0.01113&32&1.598e-09&3.732e-09&0.01312\\
64&5.111&8.449&0.01423&64&1.991e-13&5.801e-13&0.01731\\
\hline
\end{tabular}
\end{table}


\begin{table}
\caption{Comparison of the COS and SFP methods in error convergence and CPU time for pricing European cash-or-nothing put options under the BSM model. The range of $K$ is from 80 to 120. Spectral convergence is observed in the COS and SFP methods. The parameters are taken from \textbf{BSM--Para1}.} 
\label{table:BS_AllorNo_SFP}
\centering 
{\renewcommand{\arraystretch}{1.2} 
\begin{tabular}{|c|c|c|c||c|c|c|c|} 
\hline
\multicolumn{4}{|c||}{\textbf{COS}}&\multicolumn{4}{c|}{\textbf{SFP}}\\
\hline
N&$R_\infty$&$R_2$&Time&U&$R_\infty$&$R_2$&Time\\
\hline
8&6.395e-02&7.711e-01&0.00716&8&3.673e-03&3.974e-02&0.00901\\
16&2.562e-03&2.831e-02&0.00831&16&3.801e-06&2.646e-05&0.00931\\
32&7.681e-08&8.794e-07&0.01123&32&5.702e-12&2.668e-11&0.01342\\
64&1.772e-15&4.015e-15&0.01523&64&1.156e-14&2.297e-14&0.01791\\
\hline
\end{tabular}
}
\end{table}

\begin{table} 
\caption{Comparison of the COS, SWIFT and SFP methods in terms of absolute error for pricing a call option under the BSM model. The reference values have been computed using the Black-Scholes analytical formulae: 99.2025928525532000 ($T = 50$) and 99.9945609694213000 ($T=100$). The SFP method is more accurate than the others with fewer summation terms required. The parameters are taken from \textbf{BSM--Para3}.} 
\label{table:BS_LongMa_SFP}
\centering 
{\renewcommand{\arraystretch}{1.2} 
\begin{tabular}{|ccc|} 
\hline
Method&Error $(T=50)$& Error $(T=100)$\\
\hline
SWIFT $(m = 0)$ &$1.91e-01$&$2.50e-05$\\
COS $(N = 35)$ &$4.98e-01$&$2.05e+02$\\
SFP $(U=32)$&$2.653e-07$&$7.067e-08$\\
SWIFT $(m = 1)$& $7.78e-09$& $3.20e-06$\\
COS $(N = 70)$& $2.79e-08$ &$2.02e-05$\\
SFP $(U=64)$&$2.251e-10$&$7.037e-11$\\
\hline
\end{tabular}}
\end{table}
\begin{figure}
\center
\includegraphics[height=6cm,width=10cm]{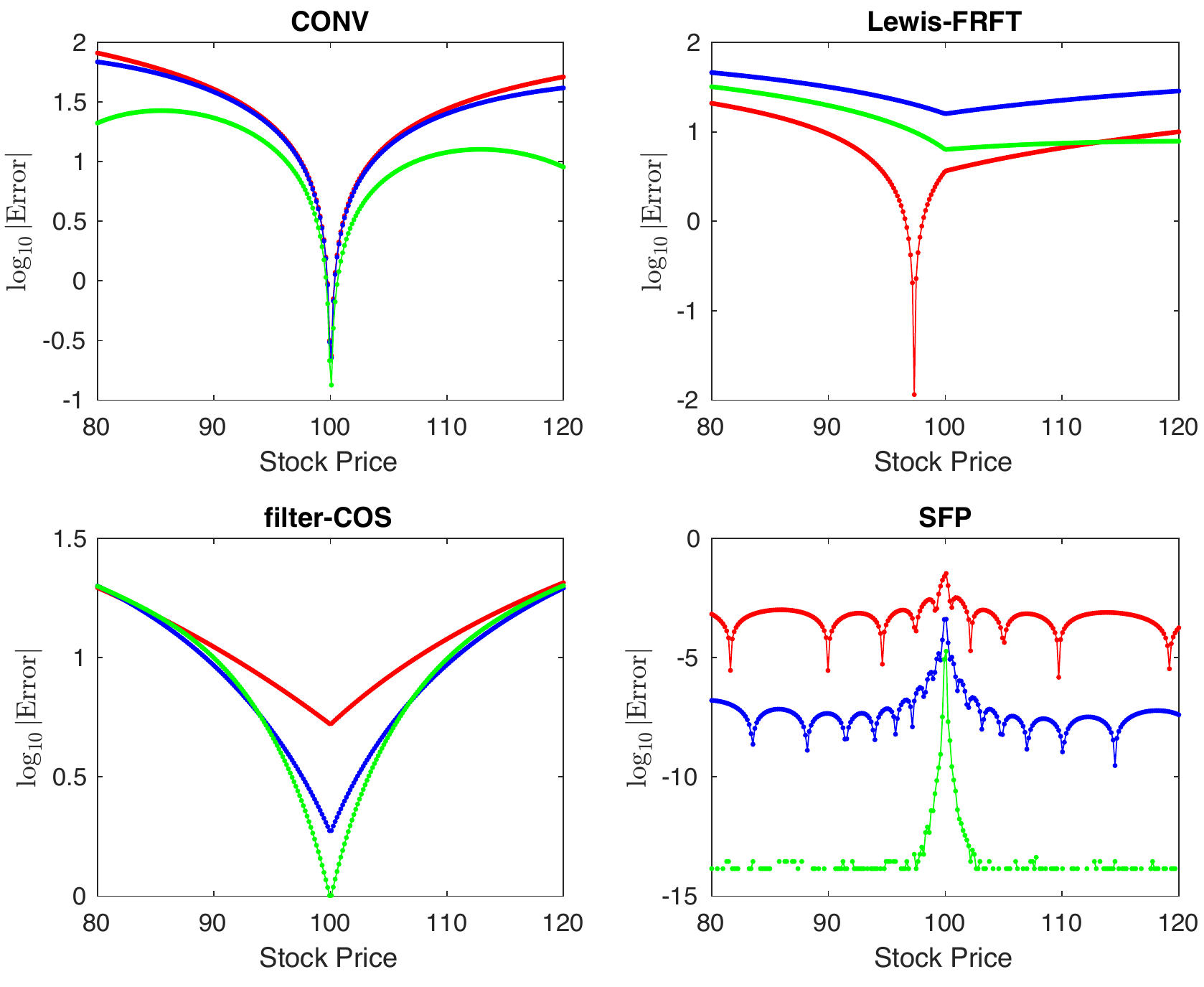}
\setlength{\abovecaptionskip}{1pt}
\caption{Error envelopes of the CONV, Lewis-FRFT, filter-COS and SFP methods when pricing European at/around-the-money call options under the BSM model in an $S_0$ range from 80 to 120. The number of terms/grid points increases in a sequence of  8 (red line), 16 (blue line) and 32 (green line). In the SFP method, spectral convergence is observed away from the jump. The parameters are taken from \textbf{BSM--Para4}.}
\label{fig:BS_shortT_SFP}
\end{figure}

\begin{table}
\caption{Comparison of the CONV, COS, filter-COS and SFP methods in terms of the absolute error for pricing a call option under the BSM model.  The reference values are  $0$ ($S_0= 95$) and $0.007491657716010$ ($S_0=99.999$). The SFP method is more accurate than the others with fewer summation terms required. The parameters are taken from \textbf{BSM--Para4}.} 
\label{table:BS_Sing_SFP_individuals}
\centering 
\resizebox{\textwidth}{!}{
\begin{tabular}{|cccc||cccc|} 
\hline
$S_0= 95$&\textbf{CONV}&\textbf{filter-COS}&\textbf{SFP}&$S_0=99.999$&\textbf{COS}&\textbf{filter-COS}&\textbf{SFP}\\
\hline
$N/U$&  Error &Error &Error &$N/U$&  Error &Error &Error \\
\hline
8&    2.004e+01&    5.099&    2.813e-04& 8&   2.475e-01&   4.482e-01&   3.849e-02\\
16&    1.940e+01&    4.987&    7.268e-08&16&   1.307e-01&   1.944e-01&   4.704e-03\\
32&   1.311e+01&    5.001&    0.000& 32&   7.008e-02&  1.030e-01&  3.473e-03\\
64&    6.696 &         5.000&    0.000&   64& 3.960e-2&   5.587e-02&   6.268e-05\\
\hline
\end{tabular}
}
\end{table}

\begin{figure}
\center
\includegraphics[height=6cm,width=10cm]{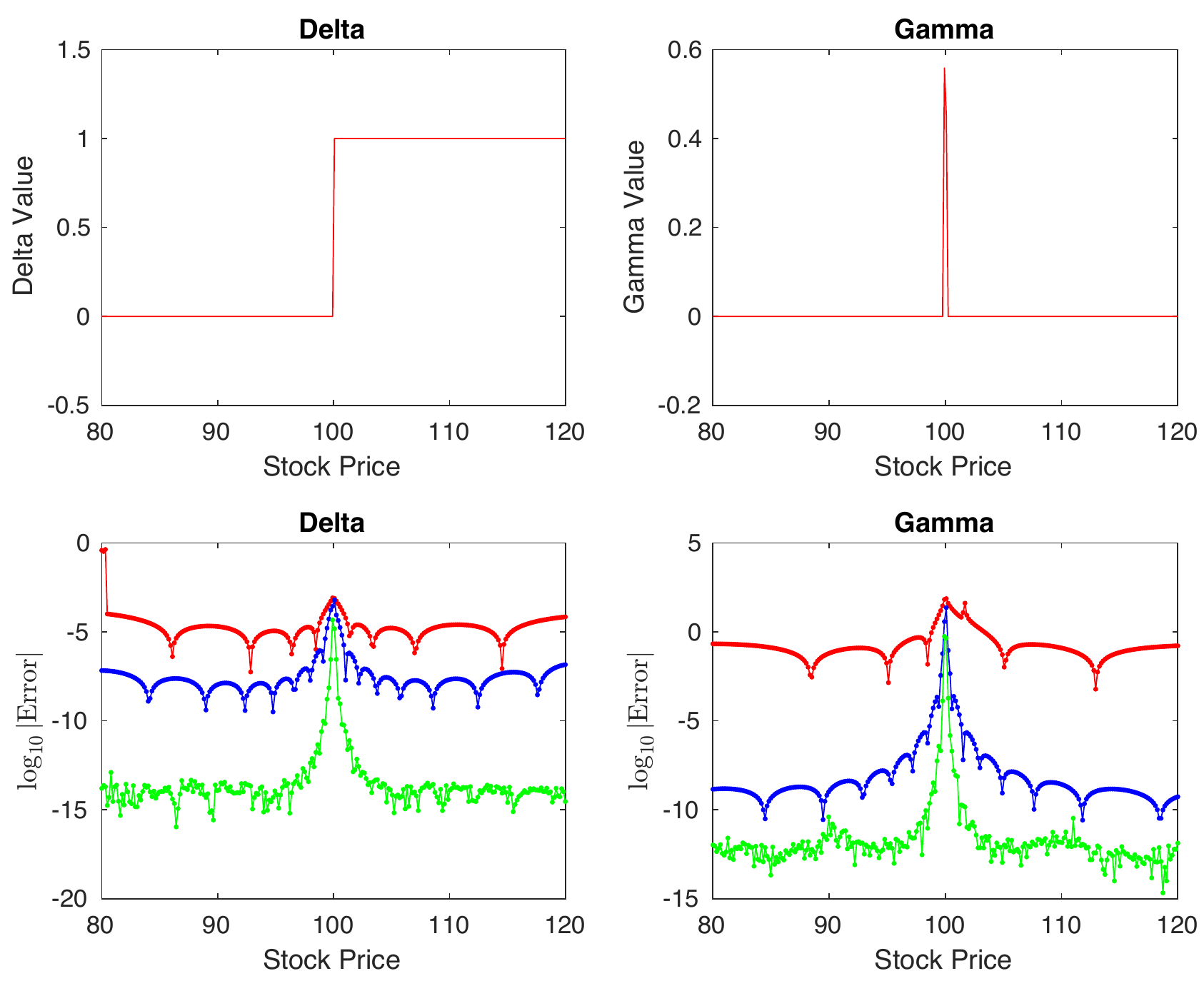}
\setlength{\abovecaptionskip}{1pt}
\caption{Recovered call Delta $\Delta$ (top left) and Gamma $\Gamma$ (top right) via the SFP method under the BSM model with strikes increasing from 80 to 120 and their corresponding error envelopes: Delta $\Delta$ (bottom left) and Gamma $\Gamma$ (bottom right). The number of terms increases in a sequence of  8 (red line), 16 (blue line) and 32 (green line). In the SFP method, spectral convergence is observed away from the jump. The parameters are taken from \textbf{BSM--Para4}.}
\label{fig:BSDeltaGammaSFP}
\end{figure}

\subsubsection{The Variance Gamma Model}\label{subsection:VG}
A VG process \citep{Mad_Sen:1990, Mad_Mil:1991,Car_Mad_Cha:1998} is an infinite activity L\'evy process and is a subordinate version of Brownian motion \citep[cf.][]{Con_Tan:2004}. 
The most important feature of this model is that the Brownian motion is evaluated in \textsl{random time} $t^{*}$ (determined by an independent increasing L\'evy process---a Gamma process) rather than in calendar time $t$. Suppose that the VG process $b(t^{*};\theta,\sigma)$ is defined as $\theta t^{*}+\sigma W_{t^*},$ where the random time $t^{*}$ is given by a Gamma process $\rm Gamma(t;1,\upsilon)$ with a unit mean and variance $\upsilon$, $\theta$ is a drift at $t^{*}$, and $W_{t^*}$ denotes a standard Brownian motion. Then, we define the stock price dynamics driven by the VG process as follows:
\begin{align}
S_T=S_0e^{(r-q-\omega)t+\theta {\rm Gamma(t;1,\upsilon)}+ \sigma W_{{\rm Gamma(t;1,\upsilon)}}},
\end{align}
and its characteristic function is given by 
\begin{align}
\varphi(u)=\exp\bigg(iu(r-q-\omega)t\bigg) \left(\frac{1}{1-i\theta\upsilon u+\frac{\sigma^2\upsilon}{2}u^2}\right)^\frac{t}{\upsilon}, \,\, u\in\mathbb{R}.
\end{align}
Here $\omega=-{1\over \nu}\log\left(1-\theta\upsilon-\frac{\sigma^2\upsilon}{2}\right).$ 
The parameters for the experiments are selected from the followings:
\begin{align}
\textbf{VG--Para1}:\, S_0&=100,\, \sigma=0.12,\,  \theta=-0.14,\,\nu=0.2,\, r=0.1,\, T=0.1,\, q=0,\\ 
\textbf{VG--Para2}:\, K&=1,\, \sigma=0.1213,\,  \theta=-0.1436,\,\nu=0.1686,\, r=0.03,\, T=1,\, q=0.01.
\end{align}

The first set of parameters is chosen because relatively slow convergence was reported for the CONV method for very short maturities in \cite{Lor_Fan:2008}. The last set is from \cite{Pis_Sto:2012}, which originates from \cite{Car_Mad_Cha:1998}. The reference values for these tests are based on the VG analytical formula. We present all the recovered PDFs via the SFP method in Figure \ref{fig:VG_density_SFP}, and the non-smooth recovered PDF (red line) can be obtained via \textbf{VG--Para1}. We set $L=10$ in the truncated intervals of (\ref{eqn:truncInt1}) and (\ref{eqn:truncInt2}). In first numerical test (\textbf{VG--Para1}), we check for convergence behaviour against 250 different call prices within a range of $K$ from $80$ to $120$ and then examine some individual call prices around/at the jump. Global spectral convergence away from the jump is reported for the SFP method in Figure \ref{fig:VG_Sing_SFP}. In Table \ref{table:VG_Sing_SFP_individuals}, when $K=90,$ the SFP method can achieve 13 digits of accuracy with 128 terms required. This result is far better than those of the other two methods, as they only achieve 4 or 5 digits of accuracy. In the same test, \cite{Fan_Oos:2009a} show that the COS method requires 1024 terms to achieve an error difference of $2.52e-08$ from the true solution. Obviously, their result indicates that the COS method cannot compete with the SFP method. When $S_0=102.336,$  the call price is measured at the jump. We can see that the convergence rate of the call price becomes algebraic. Compared with other methods, the SFP method is still more accurate with 6 digits of accuracy and only 128 terms required. This is in line with the finding in \cite{Dris_Ben:2011} that the SFP method can yield 4-6 digits of accuracy at the jump if the function is very difficult. In the last test (\textbf{VG--Para2}), we check for convergence behaviour against 250 call prices within a range of $S_0$ from $0.5$ to $2.$ Figure \ref{fig:VG_Smooth_SFP} and its numerical representation---Table \ref{table:VG_Smooth_SFP1}---suggest that the COS and SFP method have global spectral convergence when the PDF is smooth. However, the SFP method has the highest accuracy in terms of $R_\infty$ and $R_2.$ Moreover, in the test, we notice that the COS method yields less stable and accurate numerical results when it is used to approximate small values of option prices.

\begin{figure}
\center
\includegraphics[height=5cm,width=7cm]{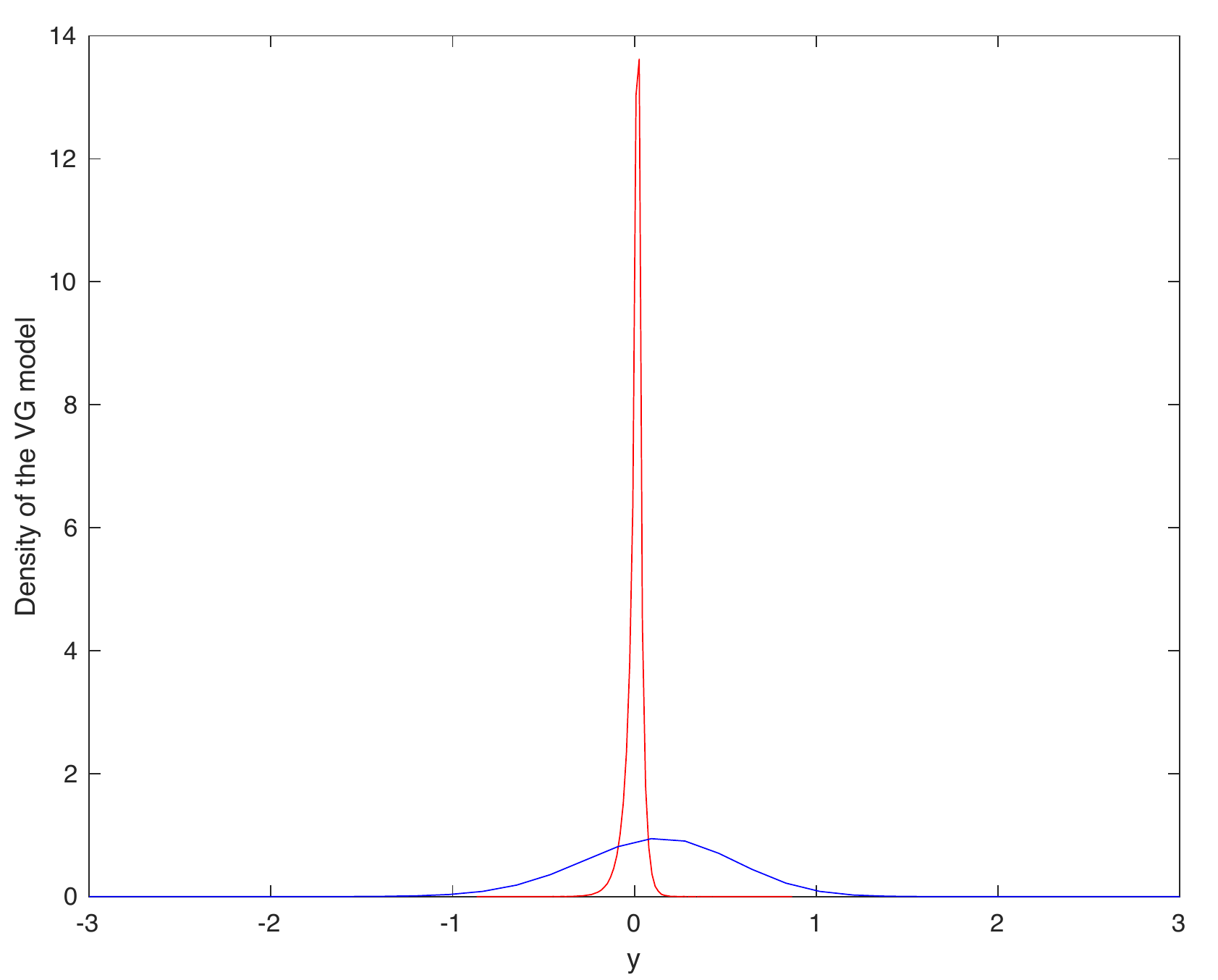}
\setlength{\abovecaptionskip}{1pt}
\caption{Recovered density functions for the VG model are generated via two sets of parameters: \textbf{VG--Para1} (red line) and \textbf{VG--Para2} (blue line).}
\label{fig:VG_density_SFP}
\end{figure}

\begin{figure}
\center
\includegraphics[height=6cm,width=10cm]{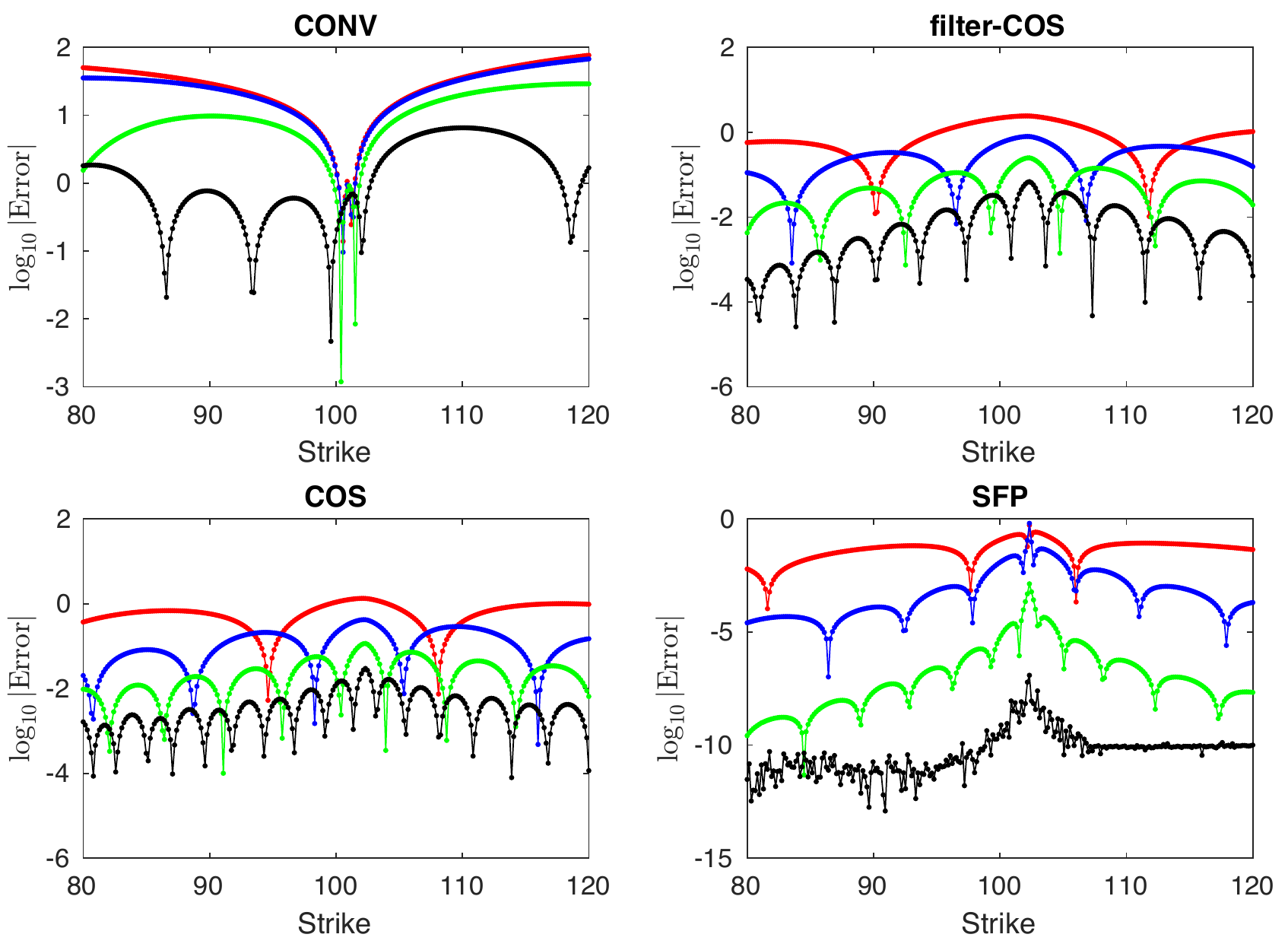}
\setlength{\abovecaptionskip}{1pt}
\caption{Error envelopes of the CONV, filter-COS, COS and SFP methods for pricing European at/around-the-money call options under the VG model. The strikes increase from 80 to 120. The number of terms/grid points increases in a sequence of  8 (red line), 16 (blue line), 32 (green line) and 64 (black line). Apart from the jump, spectral convergence is observed in the SFP method. The parameters are taken from \textbf{VG--Para1}.}
\label{fig:VG_Sing_SFP}
\end{figure}

\begin{table}
 \caption{Comparison of the COS, filter-COS and SFP methods in terms of absolute error for pricing a vanilla call option under the VG model.  The reference values are  $10.993703186728190$ ($K=90$) and $0.689027011772653$ ($K = 102.336$). The SFP method is more accurate than the others with fewer summation terms required. The parameters are taken from \textbf{VG--Para1}.} \label{table:VG_Sing_SFP_individuals}
\centering 
\resizebox{\textwidth}{!}{\begin{tabular}{|cccc||cccc|} 
\hline
$K=90$ &\textbf{COS}&\textbf{filter-COS}&\textbf{SFP}&$K=102.336$&\textbf{COS}&\textbf{filter-COS}&\textbf{SFP}\\
\hline
$N/U$&  Error &Error &Error &$N/U$&  Error &Error &Error \\
\hline
8&5.783e-01&3.062e-02&5.256e-02&8&1.326&2.413&5.506\\
16&5.843e-02&3.152e-01&1.241e-04&16&4.185e-01&7.921e-01&5.435\\
32&1.405e-02&4.756e-02&2.221e-08&32&1.153e-01&2.488e-01&2.895e-04\\
64&1.603e-03&7.951e-04&1.401e-11&64&2.976e-02&6.836e-02&2.899e-05\\
128&4.281e-04&8.931e-05&5.755e-13&128&7.595e-03&1.762e-02&1.147e-06\\
\hline
\end{tabular}}
\end{table}

\begin{figure}
\center
\includegraphics[height=6cm,width=10cm]{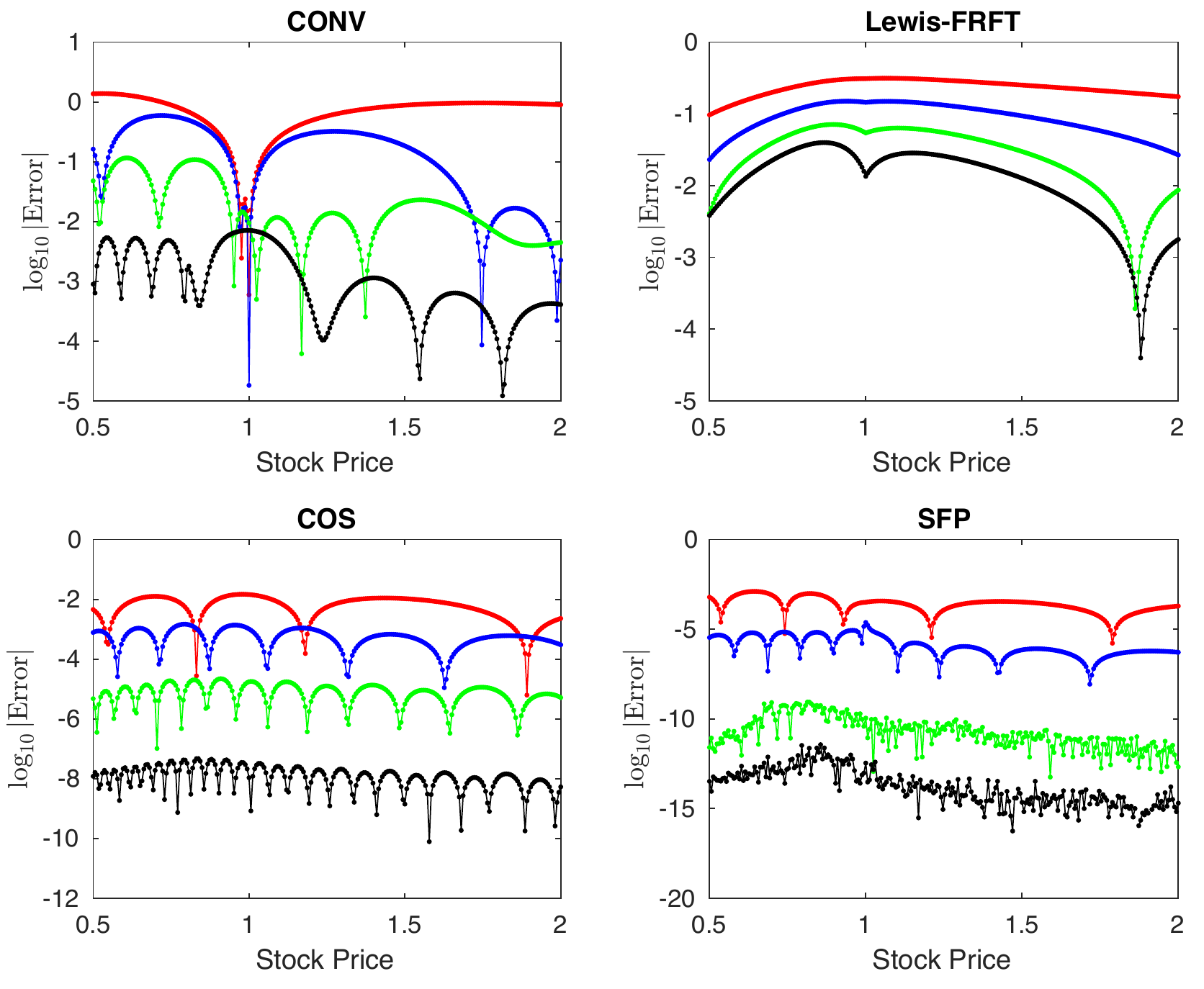}
\setlength{\abovecaptionskip}{1pt}
\caption{Error envelopes of the CONV, Lewis-FRFT, COS and SFP methods for pricing European at/around-the-money call options under the VG model in the range of $S_0$ from 0.5 to 2. The number of terms/grid points increases in a sequence of  8 (red line), 16 (blue line), 32 (green line) and 64 (black line). Spectral convergence is observed for the COS and SFP methods. The SFP method has the highest accuracy. The parameters are taken from \textbf{VG--Para2}.}
\label{fig:VG_Smooth_SFP}
\end{figure}

\begin{table}
 \caption{Comparison of the CONV, Lewis-FRFT, COS and SFP methods in terms of error convergence and CPU time for pricing European at/around-the-money call options under the VG model in the range of $S_0$ from 0.5 to 2. Spectral convergence is observed in both the COS and SFP methods. The SFP method has the highest accuracy. The parameters are taken from \textbf{VG--Para2}.} 
\label{table:VG_Smooth_SFP1}
\centering 
\begin{tabular}{|c|cc|c||c|cc|c|} 
\hline
\multicolumn{4}{|c||}{\textbf{CONV}}&\multicolumn{4}{c|}{\textbf{Lewis-FRFT}}\\
\hline
N&$R_\infty$&$R_2$&Time&N&$R_\infty$&$R_2$&Time\\
8&1.384&13.31&0.00816& 8&3.147e-01&3.913&0.02111\\
16&5.933e-01&4.344&0.00941&16&1.506e-01&1.645&0.02625\\
32&1.161e-01&6.922e-01&0.01012&32&7.126e-02&6.727e-01&0.04571\\
64&7.092e-03&4.533e-02&0.01641&64&3.986e-02&3.245e-01&0.09151\\
\hline
\multicolumn{6}{c}{\textbf{}}\\
\hline
\multicolumn{4}{|c||}{\textbf{COS}}&\multicolumn{4}{c|}{\textbf{SFP}}\\
\hline
N&$R_\infty$&$R_2$&Time&N&$R_\infty$&$R_2$&Time\\
8&1.477e-02&1.292e-0&0.00716&8&2.174e-03&1.476e-02&0.00921\\
16&1.472e-03&1.131e-02&0.00841&16&5.234e-05&2.300e-04&0.00961\\
32&2.217e-05&1.596e-04&0.01223&32&2.409e-09&7.566e-09&0.01372\\
64&4.694e-08&2.857e-07&0.01413&64&1.541e-11&2.485e-11&0.01741\\
\hline
\end{tabular}
\end{table}

\subsubsection{The CGMY Model}\label{subsection:CGMY}
The CGMY model was developed by \cite{Car_Ger_Mad_Yor:2002} and can be seen as a generalisation of the VG model discussed above. \citet{Car_Ger_Mad_Yor:2002} introduced the CGMY model as a class of infinitely divisible distributions (also known as tempered stable processes; see \citealp{Con_Tan:2004}). The L\'evy measure for the CGMY process is given by
\begin{eqnarray}\label{Chptr2:eqn:CGMY_density function}
\nu_{\rm (CGMY)}(d\chi)=\begin{cases}&C\exp(-G |\chi|)/|\chi|^{Y+1}\mathrm{d} \chi, \,\, \chi<0, \\&C\exp(-M |\chi|)/|\chi|^{Y+1}\mathrm{d}\chi, \,\, \chi>0,\end{cases}
\end{eqnarray}
where $C>0 $, $G>0$, $M>0$, and $Y<2$. The parameter $Y$ captures the fine structure of the process. For $Y<-1$, we obtain a compound Poisson process that has finite variation and finite activity. However, when $Y\in [0,1)$, the process has infinite activity and finite variation, which is similar to a VG process (we can see that when $Y=0$, this process is equivalent to a VG process). For $Y\in [1,2)$, the process has infinite activity and infinite variation. In this paper, we focus on a CGMY process with $Y\in (0,2)/\{1\}$, so its characteristic function is defined as follows:
\begin{align}\label{eqn:CGMY_charexponent_Y2}
\varphi(u)&=\exp\Bigg(iu(r-q+\omega)+C\Gamma(-Y)G^Y\left(\left(1+\frac{iu}{G}\right)^Y-1-\frac{iuY}{G}\right)\nonumber\\
               &\,+C\Gamma(-Y)M^Y\left(\left(1-\frac{iu}{M}\right)^Y-1+\frac{iuY}{M}\right)\Bigg).
\end{align}

Here, $\omega=\varphi(-i).$ The parameters (\textbf{CGMY--Para1}) for the numerical test are drawn from \cite{Fan_Oos:2009a} and defined as follows:
\begin{align}
\textbf{CGMY--Para1}:\,& S_0=100,\, C=1,\, G=5,\, M=5,\, Y=0.5 \mbox{ or } 1.5 \mbox{ or }1.98,\\ \nonumber
                                    & r = 0.1,\, T=1,\, q=0.
\end{align}
We evaluate the SFP method's convergence rate for vanilla calls and puts under the CGMY model. \cite{Alm_Oos:2007} and \cite{Wan_For:2007} have reported that using the finite difference method to solve partial differential integral equations to obtain option prices is difficult in cases when $Y\in [1,2).$ 
Therefore, we evaluate the SFP method using $Y = 0.5,$ $Y = 1.5,$ and $Y = 1.98$ and compare the numerical results with those of the COS and CONV methods. For the cases of $Y = 0.5,$ and $Y = 1.5,$ we compute the call reference values for the numerical experiments using the COS method with $N = 2^{14}.$ However, for the case of $Y=1.98,$ we use the SFP method with $U=2^9$ to generate the put reference values. We use $L=10$ in the truncated interval of (\ref{eqn:truncInt1}), as all the input PDFs are smooth. In Figure \ref{fig:CGMY_density_SFP}, the recovered density functions for the three cases are plotted. As when $Y$ tends to 2, the tails of the PDF are fatter and heavier, and the centre of the distribution shifts. Table \ref{table:CGMY_Smooth_SFP_individuals} replicates Table 7 in \cite{Fan_Oos:2009a}. The table compares our method with the CONV and COS methods, and we see that the SFP method can achieve better accuracy than the other methods. For example, in the SFP method, when $U$ equals 32 terms, the error difference from the true values are $2.608e-08$ (when Y=0.5) and $5.060e-10$ (when Y=1.5). The results are far better than those of other two methods with the same number of terms/grids. Figure \ref{fig:CGMY_Smooth_SFP} is the graphical result for the test of $Y=1.98.$ In the test, we first generate 250 different put prices in a range of strikes from 80 to 120 and check for the error convergence against the put prices in the range. In Figure \ref{fig:CGMY_Smooth_SFP}, we can clearly see that the SFP method obtains global spectral convergence and fairly high accuracy.

\begin{figure}
\center
\includegraphics[height=5cm,width=8 cm]{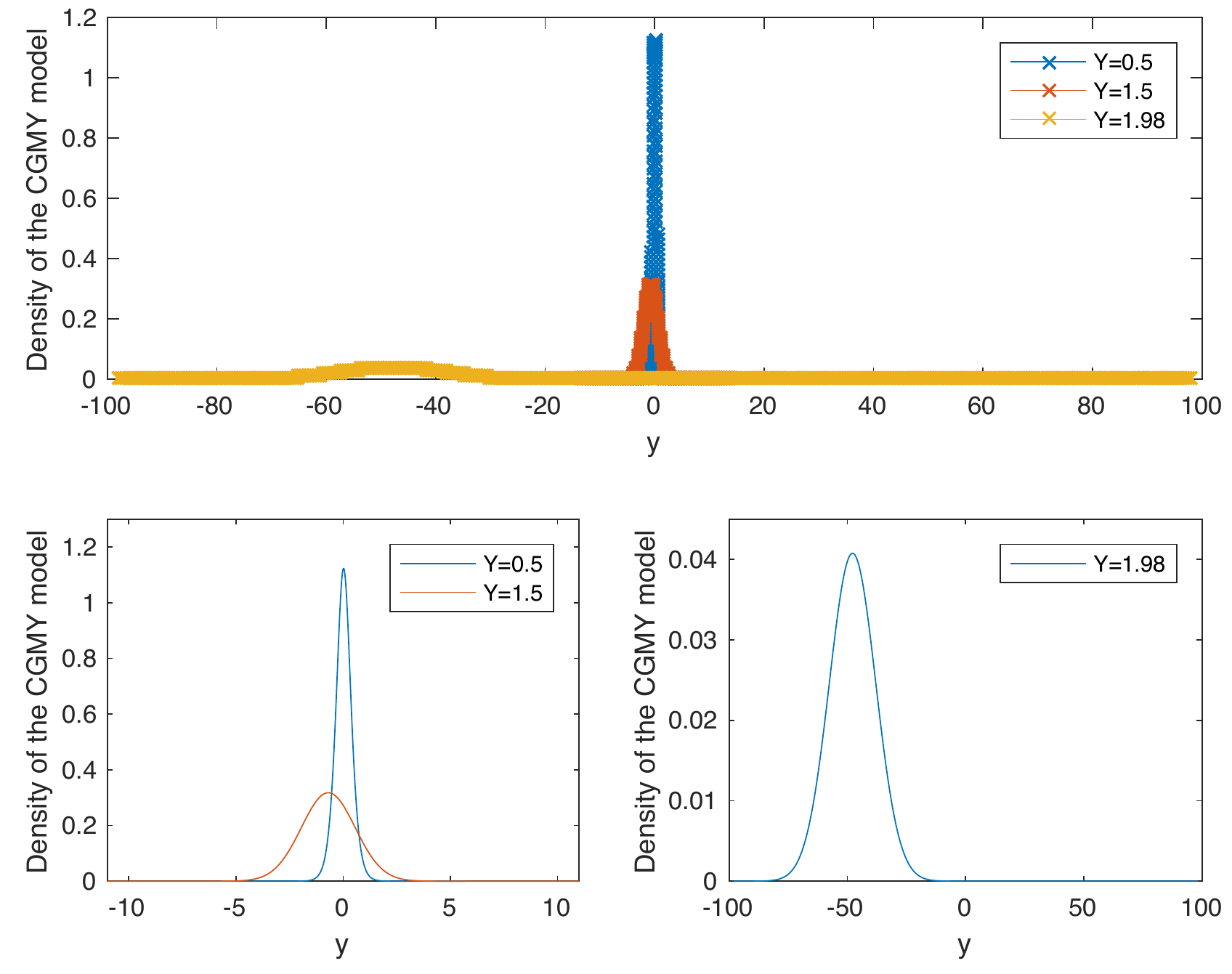}
\setlength{\abovecaptionskip}{1pt}
\caption{Recovered density functions (top) and their zoom (bottom) for the CGMY model generated via the parameters \textbf{CGMY--Para1}. }
\label{fig:CGMY_density_SFP}
\end{figure}

\begin{table}
\caption{Comparison of the CONV, COS and SFP methods in terms of absolute error for pricing vanilla call options under the CGMY model with $K=100$. The reference values are  $19.812948843118576$ ($Y= 0.5$) and $49.790905468523860$ ($Y=1.5$). The SFP method is more accurate than others with fewer summation terms required. The parameters are taken from \textbf{CGMY--Para1}.} 
\label{table:CGMY_Smooth_SFP_individuals}
\centering 
\begin{tabular}{|cccc||cccc|} 
\hline
$Y=0.5$&\textbf{CONV}&\textbf{COS}&\textbf{SFP}&$Y=1.5$&\textbf{CONV}&\textbf{COS}&\textbf{SFP}\\
\hline
$N/U$&  Error &Error &Error &$N/U$&  Error &Error &Error \\
\hline
8&    2.320&    4.443&    2.984e-02&8&   5.246&    9.304e-01&    3.020e-02\\
16&    1.079&    5.283e-01&    3.163e-04&16&    7.763e-01&    2.863e-02&    2.566e-05\\
32&    8.172e-01&    1.240e-02 &   2.608e-08&32&    7.607e-01&    1.240e-05&    5.060e-10\\
64&    2.089e-01&    2.801e-05 &   7.687e-11&48&    1.384&    5.286e-12&    8.527e-14\\
\hline
\end{tabular}
\end{table}

\begin{figure}
\center
\includegraphics[height=5cm,width=7cm]{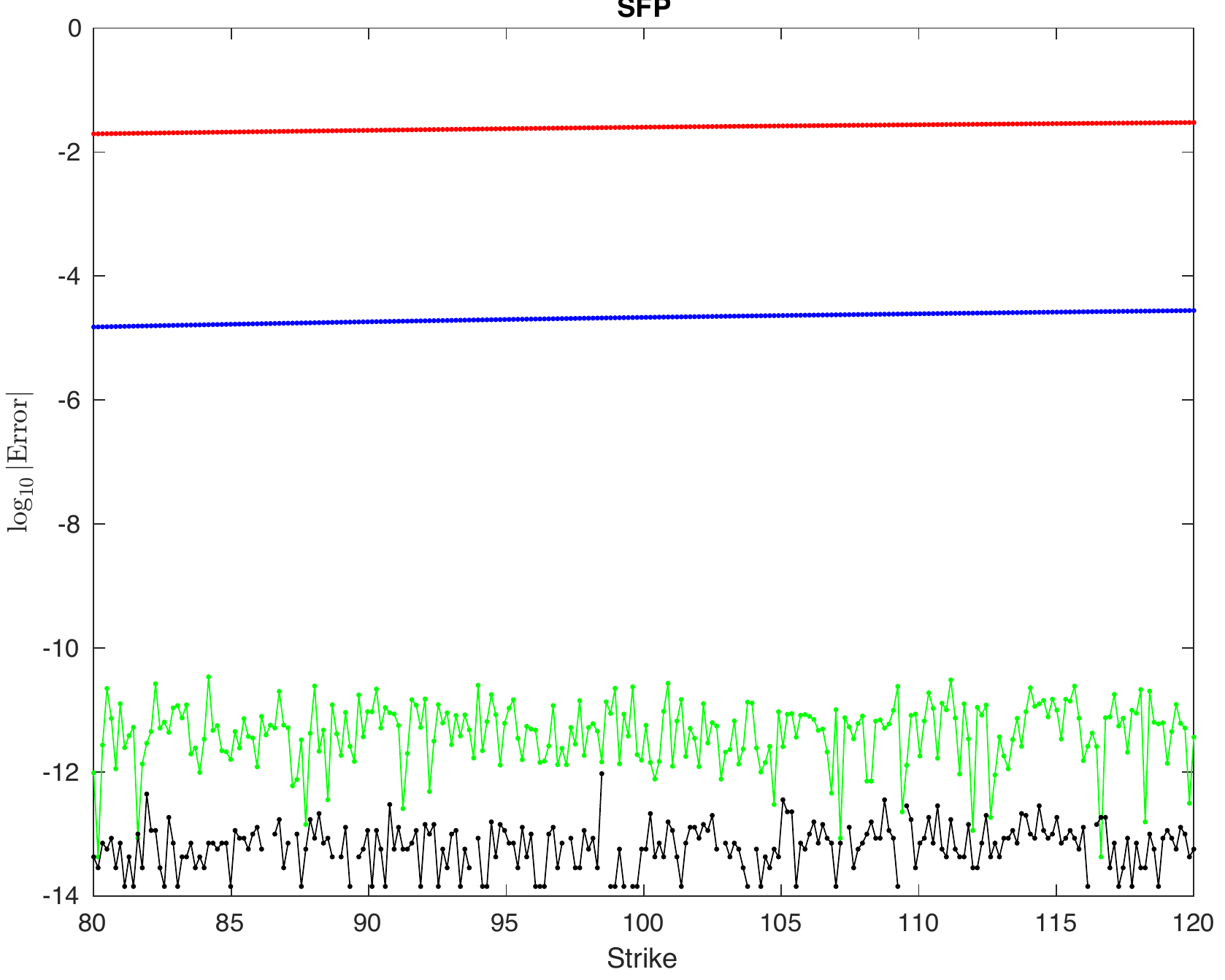}
\setlength{\abovecaptionskip}{1pt}
\caption{Error envelopes of the SFP methods for pricing European at/around-the-money put options under the CGMY model in a range of $K$ from 80 to 120 when $Y=1.98.$ The number of terms increases in a sequence of  8 (red line), 16 (blue line), 32 (green line) and 64 (black line). Spectral convergence is observed in the SFP method. The parameters are taken from \textbf{CGMY--Para1}.}
\label{fig:CGMY_Smooth_SFP}
\end{figure}


\subsection{Affine Processes}
Finally, for affine processes, we choose the Heston model and price calls using the following parameters:
\begin{align}
\small
\textbf{Heston--Para1}:& S=100,\, y_0=0.0175,\, \overline{y}=0.0398,\, \lambda=1.5768,\nonumber\\
				   &\eta=0.5751,\, \rho=-0.5711, \, r=0,\, T=1 \mbox{ or } 10 \mbox{ or } 30 \mbox{ or } 45,\, q=0.
\end{align}
Figure \ref{fig:Heston_density_SFP} presents the recovered density functions. It shows that $T = 1$ gives rise to a sharper peaked density than $T = 10,$ as expected. In this test, we compare the SFP method with the COS method. Most of the reference values are generated via the COS method with $N=2^{14}.$ However, when the option value is measured at the jump in the PDF, we generate the value using the SFP method, as the reference value provided by other methods is not accurate. We use $L=12$ in the truncated intervals of (\ref{eqn:truncInt1}) and (\ref{eqn:truncInt2}). We replicate the numerical tests shown in Tables 4 and 5 of \cite{Fan_Oos:2009a} and display them in Table \ref{table:Heston_Sing_SFP_individuals}. In the original test, \cite{Fan_Oos:2009a} (see Appendix \ref{section:Heston_OosFan}, Table \ref{table:Heston_OosFan}) report that the COS method has an algebraic convergence rate in $T=1$ and spectral convergence in $T=10$ when they approximate the call values with a strike of 100. They also suggest that the error convergence is very reasonable, e.g., the error difference is $3.17e-07$ with $N=192$ in the test of $T=1$ and $1.85e-10$ with $N=160$ in the test of $T=10.$ However, this seems not to be the case when we replicate these tests. Using \textbf{Heston--Para1} and the idea of the first and second cumulants to construct the truncated intervals (as suggested in \citealp{Fan_Oos:2009a}) for the Heston model, the PDF range is fairly large at $[-11, 11].$ According to \cite{Fan_Oos:2009a}, this implies that more terms should be used to compensate for the error convergence when the truncated interval is larger. It is not clear how \cite{Fan_Oos:2009a} obtain the results in Table \ref{table:Heston_OosFan} without requiring many terms, especially for the test of $T=1$.

Without making any changes, we follow the steps reported in \cite{Fan_Oos:2009a}  and \cite{Ruj_Oos:2013} to compute the call prices via the COS method and the filer-COS method, respectively, and compare their error convergence results with those of the SFP method. In Table \ref{table:Heston_SFP_individuals100}, the SFP method provides roughly 8 digits of accuracy in the case of $T=1$ and 10 digits of accuracy in the case of $T=10$ with only 128 terms required. This is an immense improvement over the other methods, which provide only 1-2 digits of accuracy with the same number of terms. Furthermore, if we compare our results with those shown in Tables 4 and 5 of \cite{Fan_Oos:2009a} (see Appendix \ref{section:Heston_OosFan}, Table \ref{table:Heston_OosFan}), our method yields better convergence. As the PDF is non-smooth when $T=1,$ we compare the COS, filter-COS and SFP methods in terms of absolute error for pricing call options around/at the jump. When $K=50,$ the SFP method has 14 digits of accuracy, but the other two methods gain only 4 digits of accuracy with 256 terms required. When $K=105.453,$ the option price is at the jump. The call price generated by the SFP can reach 4 digits of accuracy. This is in line with the finding of \cite{Dris_Ben:2011}. Finally,  the last test replicates Table 3 in \cite{Gra_Oos:2013}. In the test, the authors compare the error convergence of the Haar wavelets method and the COS method when they price call options with long maturities. They selected maturities $T = 30$ and $T = 45$, which may correspond to pensions or mortgage contracts. Table \ref{table:Heston_LongMa_SFP} lists the error convergence results from Table 3 in \cite{Gra_Oos:2013} and compares them with those produced by the SFP method. As we see in the table, the SFP method yields better and faster error convergence than the other methods. For example, the absolute error of the SFP method is roughly $3.000e-06$ for both $T=30 \mbox{ and } 45$ with $U=64$, but those of other two methods are range from $2.46e-03$ to $9.68e-01$ in both cases under the same conditions.

\begin{figure}
\center
\includegraphics[height=5cm,width=8 cm]{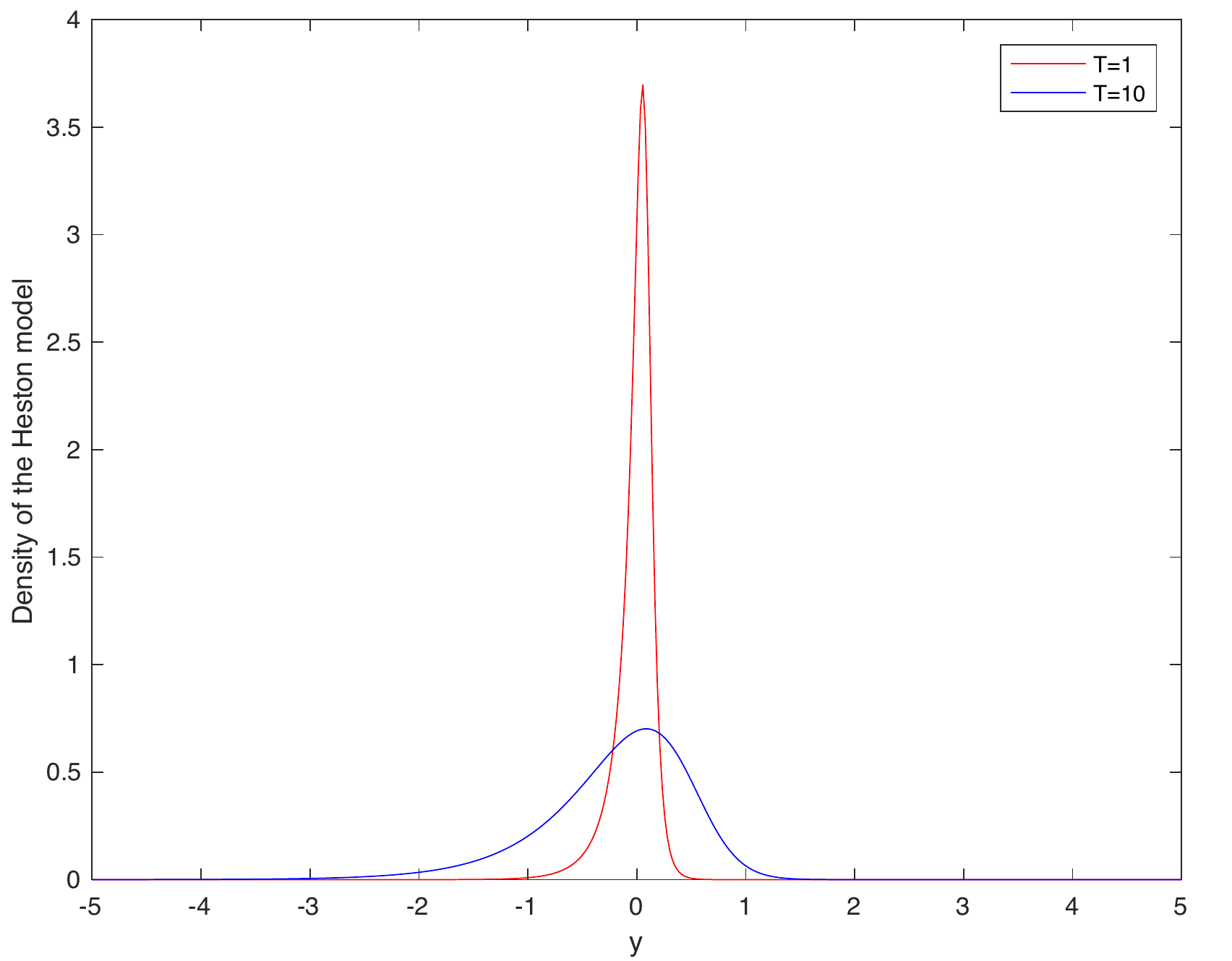}
\setlength{\abovecaptionskip}{1pt}
\caption{Recovered density functions for the Heston model generated via the parameters  \textbf{Heston--Para1}. }
\label{fig:Heston_density_SFP}
\end{figure}

\begin{table}
\caption{Comparison of the CONV, COS, filter-COS and SFP methods in terms of absolute error for pricing a call option under the Heston model with $K=100$. The reference values are  $5.7851554534076321$ ($T= 1$) and $22.318945791154533$ ($T=10$). The SFP method is more accurate than the others with fewer summation terms required. The parameters are taken from \textbf{Heston--Para1}.} \label{table:Heston_SFP_individuals100}
\centering 
\begin{tabular}{|cccc||cccc|} 
\hline
$T=1$ &\textbf{COS}&\textbf{filter-COS}&\textbf{SFP}&$T=10$&\textbf{CONV}&\textbf{COS}&\textbf{SFP}\\
\hline
$N/U$&  Error &  Error & Error &$N/U$&  Error &Error &Error \\
\hline
8&19.77&24.44&2.568&8&21.65&23.17&5.557\\
16&9.148&13.73&1.111e-01&16&15.65&18.79&4.555e-01\\
32&3.096&5.843&4.725e-02&32&7.721&12.01&2.178e-04\\
64&6.381e-01&1.752&1.262e-04&64&1.878&4.779&3.231e-05\\
128&2.685e-02&2.68e-01&1.331e-08&128&7.969e-02&7.982e-01&7.529e-10\\
\hline
\end{tabular}
\end{table}

\begin{table}
 \caption{Comparison of the COS, filter-COS and SFP methods in terms of absolute error for pricing a call option under the Heston model with $T=1$.  The reference values are  $50.070539139715081$ ($K= 50$) and $3.181555642433310$ ($K=105.453$). The SFP method is more accurate than others with fewer summation terms required. The parameters are taken from \textbf{Heston--Para1} .} \label{table:Heston_Sing_SFP_individuals}
\centering 
\resizebox{\textwidth}{!}{
\begin{tabular}{|cccc||cccc|} 
\hline
$K= 50$ &\textbf{COS}&\textbf{filter-COS}&\textbf{SFP}&$K=105.453$&\textbf{COS}&\textbf{filter-COS}&\textbf{SFP}\\
\hline
$N/U$&  Error &  Error & Error &$N/U$&  Error &Error &Error \\
\hline
8&5.031&8.546&3.514e-01&8&19.71&24.37&2.608\\
16&8.763e-01&1.202&1.047e-02&16&9.163&13.71&12.81\\
32&4.861e-01&1.197&6.327e-06&32&3.201&5.902&4.681\\
64&3.715e-02&9.661e-02&1.861e-08&64&7.801e-01&1.888&3.649e-02\\
128&6.003e-03&2.361e-02&2.467e-11&128&1.003e-01&3.941e-01&1.001e-03\\
256&3.922e-04&2.363e-04&8.527e-14&256&3.722e-03&3.801e-02&1.898e-04\\
\hline
\end{tabular}
}
\end{table}
\begin{table} 
\caption{Comparison of the COS, Haar wavelet and SFP methods in terms of absolute error for pricing a call option under the Heston model. The reference value is computed via the COS method using 50000 terms. The SFP method is more accurate than the others with fewer summation terms required. The parameters are taken from \textbf{Heston--Para1}.} 
\label{table:Heston_LongMa_SFP}
\centering 
\resizebox{\textwidth}{!}{
\begin{tabular}{|cc|cc|cc||cc|cc|cc|} 
\hline
\multicolumn{6}{|c||}{\textbf{T=30}}&\multicolumn{6}{c|}{\textbf{T=45}}\\
\hline
\multicolumn{2}{|c|}{\textbf{Haar}}&\multicolumn{2}{c|}{\textbf{COS}}&\multicolumn{2}{c||}{\textbf{SFP}}&\multicolumn{2}{c|}{\textbf{Haar}}&\multicolumn{2}{c|}{\textbf{COS}}&\multicolumn{2}{c|}{\textbf{SFP}}\\
\hline
scale&error&N&error&U&error&scale&error&N&error&U&error\\
\hline
3&2.58e+02& 8& 1.72e+06 &8&3.783&3& 3.92e+02& 8& 3.19e+07&8&2.584\\
4& 2.72e+00& 16& 2.75e+05&16&2.259e-01&4& 1.09e+01& 16& 5.45e+06&16&2.953e-01\\
5& 7.94e-01& 32& 2.19e+03&32&3.981e-04&5& 5.99e-01& 32& 3.10e+04&32&9.438e-04\\
6& 2.46e-03& 64& 3.37e-01&64&1.353e-06&6& 1.05e-02& 64& 9.68e-01&64&3.049e-06	\\
\hline
\end{tabular}
}
\end{table}

\section{Conclusion and discussion}\label{sec:conclusion}
In this paper, we illustrate how to use the SFP method \citep[cf.][]{Dris_Ben:2001, Dris_Ben:2011} as an alternative approach to hedging and pricing European-type options. We provide an error analysis to show that a good truncated interval can yield the highest accuracy of the method and theoretically prove that the approximate values generated by the method can trend towards the true option prices. 
Through the numerical experiment results, we first show that the SFP method has the ability to price any European-type option with the features of deep in/out of the money and/or very long/short maturities. Second, compared with other numerical methods of option pricing, we show that the SFP method can has faster error convergence with fewer partial summation terms required, and it remains consistently accurate when approximating large and small option prices. Third, it does not require a scale or damping factor to adjust its accuracy, e.g., as in the SWIFT, CONV and filter-COS methods. Finally, when PDFs are smooth, the method exhibits global spectral convergence. When PDFs are not smooth, it achieves global spectral convergence  except at the jumps. Nevertheless, the results indicate that the SFP method offers 4--6 digits of accuracy at the jumps. This is indeed in line with the findings of \cite{Dris_Ben:2011}.

Although the theoretical analysis/numerical results presented here have demonstrated the effectiveness of the SFP method. Further work might proceed in three ways. First, the creation of the truncated interval (\ref{eqn:truncInt2}) for a non-smooth PDF relies on a process of trial and error. In the future, a more theoretical, accurate algorithm for creating truncated intervals should be used for all non-smooth PDFs. Second, whether the SFP method can be applied to price European spread options is an interesting research question, as the SFP worked well in with one dimension (one asset). Finally,  our ultimate goal is to extend the method to price options with early exercise and/or path-dependant features.

\section*{Acknowledgement}
We thank Professor Toby Driscoll, Department of Mathematical Sciences, University of Delaware and Professor Bengt Fornberg, Department of Applied Mathematics, University of Colorado for comments that greatly improved the manuscript.

\newpage
\appendices
\section{}\label{section:cums} 
In Table \ref{table:cumulants}, we show the first $c_1,$ second $c_2,$ and fourth $c_4$ cumulants of the BSM, VG, $\rm CGMY$ and Heston models. However, as \citet{Fan_Oos:2009a} suggest, due to the lengthy representation of $c_4,$ we only present the first two cumulants of the Heston model. Given the characteristic functions, the cumulants can be generally computed using
$$c_n={1 \over i^n}{\partial^n \log \varphi(z)\over \partial z^n}\bigg\vert_{z=0}.$$ 
\begin{table}[h]
\caption{The first $c_1,$ second $c_2,$ and fourth $c_4$ cumulants of various models} 
\label{table:cumulants}
\centering 
\begin{tabular}{|l|l|} 
\hline
\multicolumn{2}{|l|}{L\'evy Models}\\
\hline 
BSM&$c_1=(r-q+\omega)t$\quad$c_2=\sigma^2 t,$\quad $c_4=0,$ $\omega=-0.5\sigma^2$\\
\hline 
VG&$c_1=(r-q+\theta+\omega)t$\\
\,&$c_2=(\sigma^2+\upsilon\theta^2)t$\\
\,&$c_4=3(\sigma^4\upsilon+2\theta^4\upsilon^3+4\sigma^2\theta^2\upsilon^2)t$\\
\,&$\omega={1 \over \upsilon}\log(1-\theta\upsilon-\sigma^2\upsilon/2)$\\
\hline 
$\rm CGMY$&$c_1=(r-q+\omega)t$\\
\,&$c_2=( C\Gamma(2-Y)(M^{Y-2} + G^{Y-2})t$\\
\,&$c_4=(C\Gamma(4-Y)(M^{Y-4} + G^{Y-4})t$\\
\,&$\omega=\left(C\Gamma(-Y)G^Y\left(\left(1+\frac{1}{G}\right)^Y-1-\frac{Y}{G}\right)+C\Gamma(-Y)M^Y\left(\left(1-\frac{1}{M}\right)^Y-1+\frac{Y}{M}\right)\right)$\\
\hline 
\multicolumn{2}{|l|}{Affine Processes}\\
\hline
Heston&$c_1=(r-q)t+(1-e^{-\lambda t}){\bar{y}-y_0\over 2\lambda }-0.5\bar{y}t$\\
\,&$c_2={1 \over 8\lambda^3}\Big(\eta t \lambda\exp(-\lambda t)(y_0-\bar{y})(8\lambda\rho -4\eta)+$\\
\,&\quad\quad$\lambda\rho\eta(1-e^{-\lambda t})(16\bar{y}-8y_0)+2\bar{y}\lambda t(-4\rho\eta+\eta^2+4\lambda^2)+$\\
\,&\quad\quad$\eta^2((\bar{u}-2u_0)e^{-2\lambda t}+\bar{y}(6e^{-\lambda t}-7)+2y_0)+8\lambda^2(y_0-\bar{y})(1-e^{-\lambda t})\Big)$\\
\hline
\end{tabular}
\end{table}
\section{Absolute error convergence for the COS method for calls under the Heston model}\label{section:Heston_OosFan} 
\begin{table}[h]
\caption{Absolute error convergence for the COS method for calls under the Heston model with $T=1$ and $T=10.$ Reference values are 5.785155450\ldots ($T=1$) and  22.318945791\ldots ($T=10$).}\label{table:Heston_OosFan}
\centering 
\begin{tabular}{|cc||cc|} 
\hline
$T=1$&&$T=10$&\\
\hline
$N$&  Error & $N$&Error\\
\hline
64& 4.92e-03&32&7.40e-03\\
96& 2.99e-04&64&5.02e-05\\
128&1.94e-05&96&1.40e-07 \\
160&2.99e-06&128&4.92e-10\\
192&3.17e-07&160&1.85e-10\\
\hline
\end{tabular}
\end{table}
\newpage
\bibliographystyle{rQUF}
\bibliography{Q4MMZ42B_SingFS}
\end{document}